\documentclass[onecolumn]{elsart3p}
\usepackage{amsmath,amssymb,amsfonts,bm}
\usepackage{lscape, graphicx}
\usepackage{subfigure}
\def\KeyWord#1{$\backslash$\IfColor{$\!\!$\textRed{#1}\textBlack}{#1}$\!\!$}

\newcommand{\be}{\begin{equation} }
\newcommand{\ee}{\end{equation} }
\newcommand{\ba}{\begin{eqnarray} }
\newcommand{\ea}{\end{eqnarray} }

\newcommand{\bmu}{\mbox{\boldmath $\mu $}}
\newcommand{\balpha}{\mbox{\boldmath $\alpha $}}

\def\half{{1\over 2}}
\def\a{{\alpha}}

\def\a{{\alpha}}

\def\bd{{\mathbf d}}

\def\bH{{\mathbf H}}
\def\br{{\mathbf r}}

\def\bS{{\mathbf S}}

\def\bR{{\mathbf R}}

\def\bk{{\mathbf k}}

\def\mbe{{\mathbf e}}
\input{epsf}
\begin{document}

\begin{frontmatter}

\title{ Flux Hamiltonians, Lie Algebras and Root Lattices With Minuscule Decorations}

\author{R. Shankar}
\address{Sloane Laboratory of Physics, Yale University, New Haven, CN 06520}
\address{Princeton Center for Theoretical Physics, Princeton University, Princeton, NJ 08544}
\author{ F. J. Burnell}
\address{Department of Physics, Princeton University, Princeton, NJ 08544}
\author{ S. L. Sondhi}
\address{ Department of Physics, Princeton University, Princeton, NJ 08544}
\address{Princeton Center for Theoretical Physics, Princeton University, Princeton, NJ 08544}

\begin{abstract}
We study a family of Hamiltonians of fermions hopping on a set of
lattices in the presence of a background gauge field. The lattices
are constructed by decorating the root lattices of various Lie
algebras with their minuscule representations. The Hamiltonians
are, in momentum space, themselves elements of the Lie algebras in
these same representations. We describe various interesting
aspects of the spectra---which exhibit a family resemblance to the
Dirac spectrum, and in many cases are able to relate them to known
facts about the relevant Lie algebras. Interestingly, various
realizable lattices such as the kagom\'{e} and pyrochlore can be
given this Lie algebraic interpretation and the particular flux
Hamiltonians arise as mean-field Hamiltonians for spin-1/2
Heisenberg models on these lattices.
\end{abstract}

\end{frontmatter}

\section{Introduction and Outline}

In this paper we study a family of Hamiltonians of fermions hopping around on
various lattices in the presence of a background gauge field. These
Hamiltonians are interesting to us, and we hope to the reader as well,
in three distinct contexts: the search for flux phases in quantum magnets,
the theory of Lie algebras, and their possessing interesting continuum limits
with a family resemblance to the Dirac Hamiltonian. Let us now expand on
these connections.

\subsection{Flux phases} \label{LargeNIntro}

The first of these contexts is where we encountered them, which is
the search for flux phases in quantum magnets. It was first noted
by Baskaran, Zou, and Anderson \cite{BA,BAZ} that a novel
mean-field theory for $SU(2)$ invariant spin-$1/2$ Hamiltonians
could be constructed by re-representing spins as fermionic
bilinears, $\bS = \bar{\psi} \vec{\mathbf{\sigma}} \psi$, and
relaxing the constraint of a unit fermionic occupation of each
site in favor of a global constraint of a half filled band. The
mean-field treatment consists of replacing the starting
Hamiltonian, quartic in the fermions, by one quadratic in them
wherein the fermions hop on the lattice in the presence of a
self-consistently calculated background (frozen) gauge field.
Following this, Affleck and Marston \cite{AffleckMarston} noted
that by extending the model to $SU(N)$ spins this mean-field
theory could be made exact at $N = \infty$ whence it could serve
as the starting point of a $1/N$ expansion. Specifically, they
showed that the Heisenberg model on the square lattice exhibited
an enticing mean-field solution with a flux of $\pi$ (a factor of
$e^{i\pi}$) per plaquette. Remarkably, they found that the
solution to this hopping problem led to a Dirac fermion at zero
energy. Since the system had particle-hole symmetry, and was at
half-filling due to its insulating parentage mentioned above, the
Dirac fermion was right at the Fermi energy and thus central to
low energy physics. In particular, fluctuations around the saddle
point were described by the Dirac fermion minimally coupled to the
fluctuating gauge field. While it is tangential to our purposes in
this paper, we note that the latter problem has been the focus of
much progress in recent years \cite{HermeleAlg}. Following these
early developments, there has been much work examining various
``flux phase'' mean-field theories on various lattices which is
too large a literature for us to review here.\footnote{See, for
example, \cite{WenPSG}, \cite{Hastings}, \cite{HermeleKagome} and
references therein for more recent work in 2 d.  A 3 d example and
connections to the quantum Hall effect were addressed by
\cite{LaughlinZou}.} Most interesting for our purposes was the
generalization to time reversal (T) breaking phases made by Wen,
Wilczek and Zee \cite{WWZ} which thus gave a mean-field meaning to
the chiral flux phases proposed previously by Kalmeyer and
Laughlin \cite{kl} on the basis of an inspired ansatz. These
phases exhibit fluxes through closed loops which are different
from the two T-invariant values $0$ and $\pi$. There is one last
aspect of this body of work that is also worth noting at the
outset, namely that it does not take the actual energetics at
$N=\infty$, and thus the relative stability of various mean-field
solutions, too seriously. Strictly at $N=\infty$ the kinds of
solutions discussed above lose out to fully dimerized mean-field
states of lower energy \cite{Dimers,ReadSachdev}. Their continuing
interest has to do with the possibility that this relative
ordering of energies is reversed as $N$ is decreased. Indeed, the
relative ordering between non-dimerized solutions could also
change as $N$ is decreased and hence one can (and people do), in
good conscience, start out by studying various interesting
mean-field solutions which at least exhibit local stability
\cite{WenPSG}.

With the above recital we can now locate our Hamiltonians: they arise, with a few caveats and
exceptions, as mean-field Hamiltonians for nearest-neighbor spin-$1/2$ Heisenberg Hamiltonians on
the appropriate lattices. They also involve, in almost all cases, T-breaking.

\subsection{Lie Algebras and Hamiltonians}

There is, of course, a large set of such Hamiltonians and we next
need to describe the restrictions that generate the family that we
study. This brings us to the second context in which our
Hamiltonians can be situated and which intrigues us most: the deep
ties our Hamiltonians have to Lie algebras. These ties are
twofold: ideas from Lie algebras are central to generating the
very lattices the fermions move on and the Hamiltonians can be
written as direct sums of pieces that are Lie algebra elements in
specified representations. Consequently we find that properties of
Lie algebras also control some striking features of the resulting
spectra. Sometimes we fully understand these connections and
sometimes we do not, though in all cases we will share what we
know with the reader.

Let us first summarize how the unit cell and the underlying
lattice have direct group theoretic significance.  A discussion of
the group theory used here can be found in \cite{Georgi,GrBook}.

Consider the unit cell.  Recall that the generators of a
(semi-simple) Lie algebras can be partitioned into a maximally
commuting Cartan subalgebra $H_{i}:\left[ i=1, \ldots r
\right]=\bH$ whose eigenvalues label the weights, and a set of
ladder operators $E_{\balpha}$ and their adjoints
$E_{\balpha}^{\dag}=E_{-\balpha}$ that act on the states to raise
(lower) the weights by $\balpha$.  The vectors $\balpha$ are
called the roots. The states within any irreducible representation
(multiplet) of a Lie algebra may therefore be visualized as a
collection of points in a space of dimension $r$, called the rank.
The coordinates of the points are the simultaneous eigenvalues of
${\bf H}$.  The roots that help us move around these points are
also vectors in the same space.  For example, in the case of the
rank-2 group $SU(3)$, (whose commuting quantum numbers are
traditionally called isospin and hypercharge in the physics
literature) the fundamental (quark) representation is an inverted
triangle, the anti-quark is a triangle, and the eight-dimensional
adjoint representation is a hexagon with two null weights at the
center.  The six nonzero roots correspond to the six corners of
the hexagon.

{\em The unit cells of the lattices we consider correspond to
certain special representations called minuscule representations.
} Starting with any one state in a minuscule representation or
multiplet, we can obtain all the others by acting with the Weyl
group, the group of reflections about the hyper-planes normal to
the roots. All weights of a minuscule representation are on the
same footing and in particular all have the same length.  Thus the
quark and anti-quark are minuscule while the adjoint
representation, with weights of both zero and non-zero length, is
not.

This unit cell is now used to decorate a lattice, which is a
subset of the root lattice $L_R$. Recall that roots, like weights,
also live in $r$ dimensions. So it is possible to choose $r$
roots, called {\em simple roots}, as a basis. A basic result of
group theory is that every root is an {\em integer} linear
combination of the simple roots.  There are of course a finite
number of them in any algebra of rank $r$. The {\em root lattice
$L_R$} is the infinite lattice formed with the same basis with but
with {\em any} integer set of coefficients. For $SU(3)$, the six
roots form a hexagon, while the root lattice is the infinite
hexagonal lattice.

{\em The lattice on which our fermions move is $L_{2R}$, the
subset of $L_R$ whose points have \underline{even} integer
coefficients, decorated by a basis corresponding to a minuscule
representation.}

\begin{figure}[htp]
\begin{center}
\subfigure[]{
 \epsfxsize=1in
 \epsffile{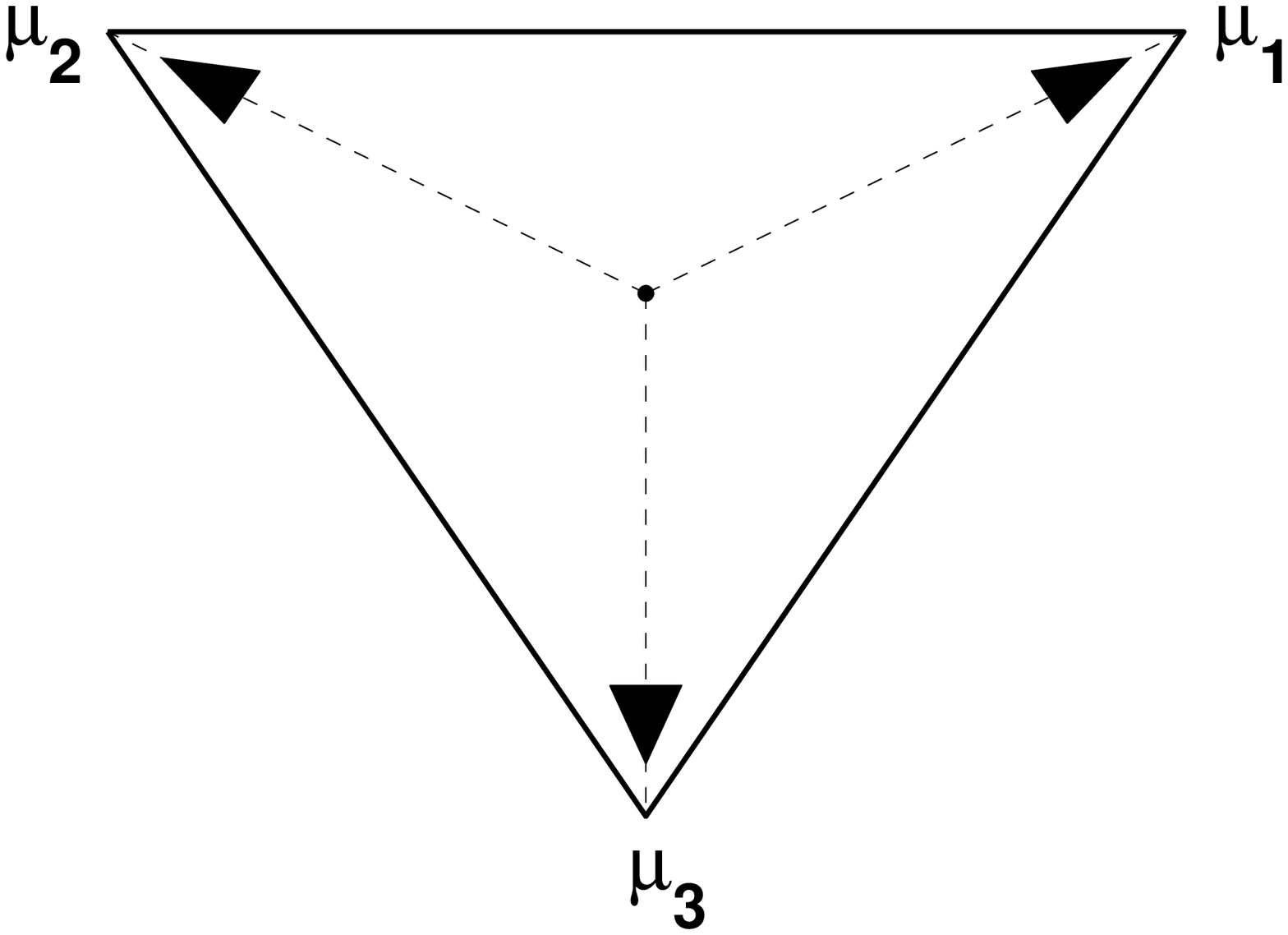} \label{KagomeUnit} }
\subfigure[]{
 \epsfxsize=2.5in
 \epsffile{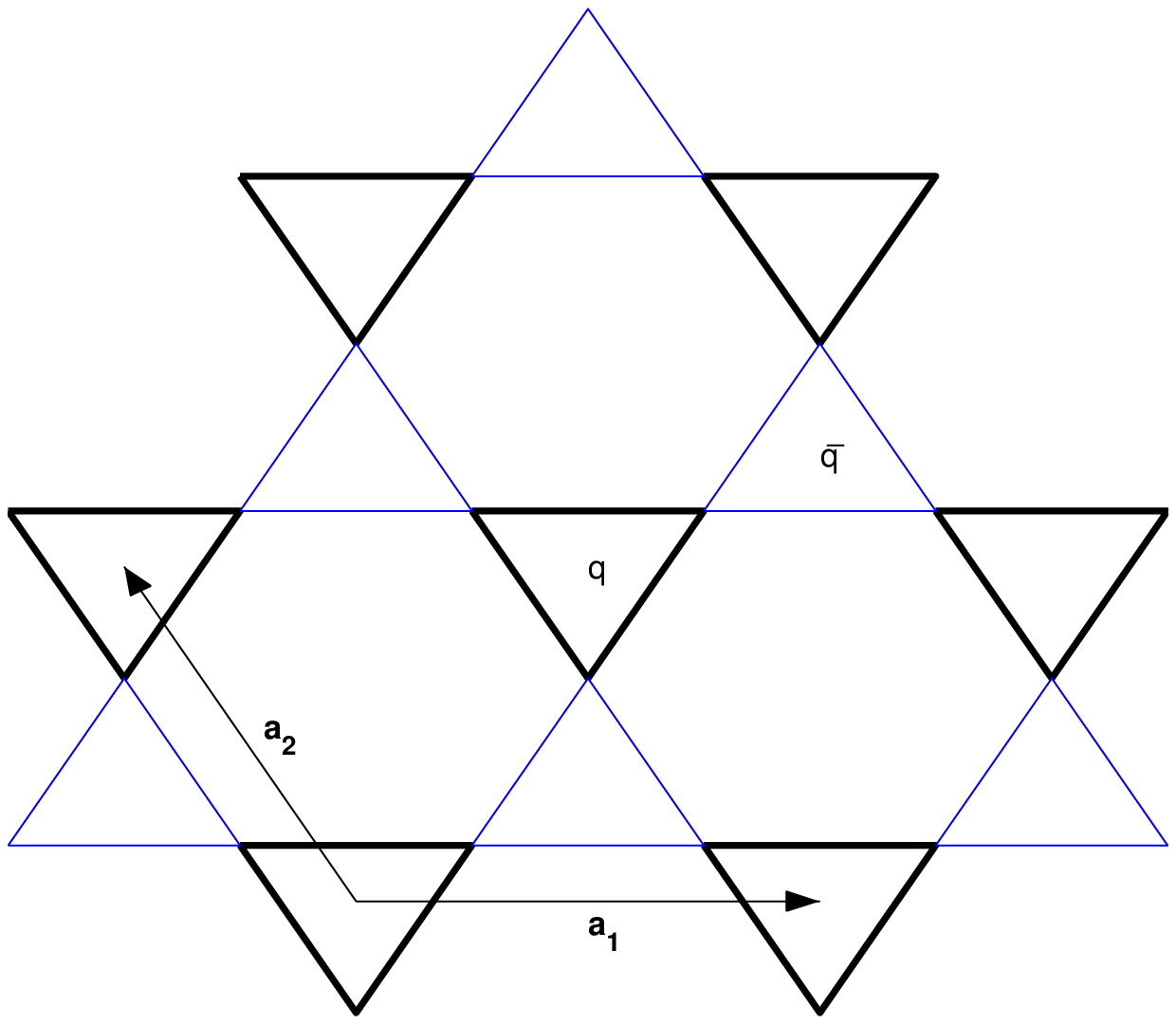}
\label{Kagome} }
 \caption{The kagom\'{e} lattice and $SU(3)$: The unit cell (a) is the quark
triplet, with weights $\mu_1 ... \mu_3$ as indicated.  The
kagom\'{e} lattice (b) is formed by decorating $L_{2R}$ (the
subset of the $SU(3)$ root lattice with even integer coefficients)
with this unit cell. $L_{2R}$ is a hexagonal lattice with basis
vectors $\mathbf{a}_1$ and $\mathbf{a}_2$ which are twice the
simple roots of $SU(3)$.  The origin of coordinates is marked $q$.
The lattice may also be viewed as being decorated by anti-quark
unit cells, one of which is centered at $\bar{q}$}.
 \end{center}
\end{figure}

When applied to the quark representation of $SU(3)$, this yields
the kagom\'{e} lattice, as shown in Fig. \ref{Kagome}. The
inverted triangle  (Fig. \ref{KagomeUnit}) is the fundamental
quark representation which forms the unit cell that decorates the
root lattice $L_{2R}$, which is a hexagonal lattice with twice the
lattice spacing as the root lattice $L_R$. The origin of
coordinates is at the point named $q$ in the figure. If you stare
at the figure hard enough, you can also see it as a hexagonal
lattice decorated by the conjugate representation, the anti-quark,
whose weights are the negatives of the quark. The center of one
such unit cell is labeled $\bar{q}$ in the figure. The quark and
anti-quark unit cells are corner sharing. These features are
common to all our models and exist because the unit cell and the
lattice are constructed from weights and roots in a particular
way. A proof of this will be given in Section \ref{LatProps}. Let
us now turn to the construction of the Hamiltonians on the above
set of lattices.  Gauge fields will enter our models in the form
of purely imaginary hopping amplitudes which can be $\pm i$. This
restriction means that on any triangular face the flux can only be
$\pm \pi /2$.\footnote{Evidently, the flux is the gauge invariant
variable.  Our choice of gauge fields is convenient for the
purposes of this paper.} In other words the background gauge field
is an Ising-like variable, and time reversal symmetry is broken.
Also, we will require that the gauge fields exhibit the
periodicity of the Bravais lattice $L_{2R}$--- this has the gauge
invariant content that there is no net flux passing through the
lattice. It is worth noting here that generically in problems of
this kind we must view symmetries as projective, i.e., the
underlying group operations will have to be accompanied by
additional gauge transformations to make the symmetry manifest
\cite{WenPSG}, say the way Lorentz transformations have to be
accompanied by gauge transformations in relativistic field
theories to establish Lorentz covariance. The classification
scheme relevant to our problem appears to be that of Color Groups,
in which each face of the crystal is colored black or white, which
we may read as $\pm \pi /2$ of flux \cite{Hammermesh,ColorGroups}.

We are now in a position to specify the Hamiltonians of interest.
Since the flux added to this lattice will be translationally
invariant on $L_{2R}$, we may go to momentum space to solve for
the dispersion relation. Evidently $H(\bk )$ will be a matrix that
acts on the states of the minuscule representation since they
constitute the unit cell. With our choices of background gauge
fields we then arrive at manifestly hermitian Hamiltonians of the
form
 \be \label{H}
 H(\bk )= \sum _{\balpha \subset {\Sigma^+}} C_{\balpha} (\bk )
(E_{\balpha} +E_{\balpha}^{\dag})
 \ee
 where the coefficients $C_{\balpha} $ are real, satisfy
  \be
  C_{\balpha}(-\bk )= -C_{\balpha}(\bk )
  \ee
and $\Sigma^+$ are the positive roots.  The roots may be divided
into positive ($\Sigma_+$) and negative ($\Sigma_-$) roots by
drawing a plane through the origin which does not contain any
roots.  This choice is basis-dependent.  For any choice of basis
we may choose an ordering of the basis vectors so that  the
positive roots are those whose first nonzero component is
positive.

Since the ladder operators move us around the multiplet, it is
reasonable to consider $H$ of the form (\ref{H}). However we must
bear in mind that this is not the most general possibility on this
lattice. For example the model only allows hops between sites that
differ by a {\em single } root while there are minuscule
representations where the states differ by more than a single
root. Other restrictions implied by this form of $H$ will be
discussed later.

Observe that we have obtained an unusual connection between the
lattice and the hopping Hamiltonian: the former is obtained by
decorating the root lattice of a Lie algebra with the weights of
one of its representations, and the hopping Hamiltonian (in
momentum space) is an element of the same Lie algebra in the same
representation! It is worth emphasizing that this connection does
not involve the symmetry group of the problem. For example, the
Lie algebra $SU(3)$ that shows up on the kagom\'{e} lattice does
not generate the actual symmetries of the hopping problem or even
correspond to any symmetries of the starting $SU(2)$ Heisenberg
model on that lattice.

One final point on the construction of our lattices and
Hamiltonians:  Although our lattice construction works in all
dimensions, we shall limit ourselves to lattices in $d=2$ and
$d=3$, both because these are the dimensions we could encounter in
the lab and because the rules for attaching flux break down in
higher dimensions wherein areas cannot be oriented unambiguously.
In this latter aspect the flux Hamiltonians that give rise to
Dirac problems are special as they involve $\pi$ flux, which does
not carry an orientation, and thus their generalization to
arbitrary dimensions is straightforward.

\subsection{Dirac-like continuum limits}

The discretized Laplace operator on lattices typically looks like
a hopping problem with zero background flux. It is an interesting
fact that when Kogut and Susskind \cite{KogutSusskind} set out to
discretize the Dirac operator on cubic lattices, they were
naturally led to introduce a background flux of $\pi$ per
plaquette. Indeed, the Affleck-Marston work ended up
``rediscovering" this earlier construction in the case of two
spatial dimensions.

The flux Hamiltonians we consider evidently have a family resemblance to the $\pi$
flux Hamiltonians which they generalize. It is also the case that their low
energy limits exhibit a family resemblance to the Dirac theory which is the third
context in which they appear to be interesting.

As a first step in explaining what we mean let us observe that
zero energy plays a special role in our entire set of
Hamiltonians. Normally, this comes about via a particle-hole
symmetry, where at every momentum $\bf{k}$ states at energy $E$
are accompanied by states at $-E$. In our problems, the choice of
background flux ensures that $H(-\bk )= -H(\bk )$. Consequently
for every level that is negative and hence occupied at $\bk$,
there is one at $-\bk$ that is empty and unoccupied so that the
combination of the two bands is particle-hole symmetric and $E=0$
is again special. We should observe that this is very useful in
the original context of the mean-field theory of various
Heisenberg models for this ensures that the half-filled band
exhibits a Fermi energy, $E_F=0$.

In the $\pi$ flux phase, the structure about $E=0$ is that of the
Dirac theory. In our examples we find generalizations that we term
Dirac-like or pseudo-Dirac. By Dirac-like, we mean a
generalization in three respects.  First, although the spectrum is
linear in momentum for small momentum, it possesses only discrete
and not full rotational invariance. Second, the square of the
Hamiltonian $H$ is not proportional to the unit matrix, though a
higher degree polynomial is. While a result of this sort is
inevitable for any finite size matrix, the fact that the
polynomial often contains just even powers leads to a result that
is sufficiently reminiscent of the Dirac case. The third
generalization we encounter is that in addition to isolated Dirac
(or Dirac-like) points, we often find entire lines and even planes
of zero energy.

The zeros are interesting in and of themselves in the Lie
algebraic setting: in some cases, the lines of zeros are along the
direction of the weights while in other cases the planes of zeros
are simply related to the roots and so on. In some cases we can
understand the locus of zeros without explicit computation by
appealing to ideas from group theory, while in many cases we could
neither anticipate nor explain the zeros.

There is, inevitably, a matrix structure that goes with the low energy
Dirac-like theory and generalizes the gamma matrices but it does not appear
to be immediately interesting in and of itself although we will exhibit
it in one especially interesting case.

\subsection{What follows}

In Section \ref{LatticeSection} we explain the construction of the
lattice in detail, and describe the examples we will study in the
remainder of the paper.  We discuss the hopping problem for the
$d=2$ lattices in Section \ref{2dFluxSection}, studying several
possible background fluxes.  In Section \ref{3dFluxSection} we
carry out the same analysis for lattices in $d=3$.  Here the
reader can find in some detail a discussion of the pyrochlore
lattice, which we understand the best in terms of group theory and
which also displays interesting mathematical structures.  In
Section \ref{CommentsSection} we comment on the utility of our
Hamiltonians as mean-field solutions of the Heisenberg model on
our various lattices and on two extensions of our analysis.
Section \ref{Conclusions} contains concluding remarks and
discusses open problems.

\section{Lattice construction} \label{LatticeSection}
 Recall our recipe for generating the lattices:
 \begin{itemize}
 \item Generate $L_{2R}$, the even sector of the root lattice, that is to
say, even integer combinations of the simple roots.
     \item Decorate each lattice point with a minuscule representation.
     \end{itemize}

As noted in the Introduction, although our lattice construction
works in all dimensions, we shall limit ourselves to lattices in
$d=2$ and $d=3$, due to their physical relevance and more
absolutely because the rules for attaching flux break down in
higher dimensions wherein areas cannot be oriented unambiguously.
Hence we study rank 2 and 3 groups only. Here is the list of
candidates.
     \begin{itemize}
     \item $d=2$: $SU(3)$ and $SO(5)$=$Sp(4)$. The exceptional group $G_2$ does
not have minuscule representations. The group SO(4) factors into
two independent $SU(2)$ factors and will only be discussed very
briefly.

     \item $d=3$: $SU(4)$=$SO(6)$, $SO(7)$, and $Sp(6)$. There are no exceptional
groups of rank 3.
         \end{itemize}
         The minuscule representations in each case will be listed as we
go along.

 \subsection{Properties of our lattices.} \label{LatProps}

The lattices we manufacture by the rules listed above have some
interesting features that will be established in this section.
Before we do so in general, let us pause to examine a simple
example, the rank-2 group $SU(3)$.  This exercise will help us
better motivate and understand the general case.

      The only minuscule representations are the quark and antiquark.
Let us begin with the quark representation.

     The weights, numbered $1, 2, 3$ in Fig. \ref{KagomeUnit}, are
     \be
     \bmu_1  =  \half (1, {1 \over \sqrt{3}}), \ \ \
     \bmu_2  =  \half (-1, {1 \over \sqrt{3}}),\ \ \
     \bmu_3  =  (0, -{1 \over \sqrt{3}})
     \ee
and point to the vertices of an equilateral triangle.

One significant feature of these weights that we will invoke is that
     \be
     \bmu_i \cdot \bmu_j =
\begin{cases}\label{sun2}
a & i=j \\
 b  & i \ne j
\end{cases}
     \ee
 where $a=1/3$ and $b=-1/6$. In other words, there are just two possible
values for $\bmu_i \cdot \bmu_j$, between a weight and itself ($a$) and
between a weight and any other ($b$).

     The nice thing about $SU(3)$  is that in the quark representation, the
weights form a simplex where every corner is equidistant from
every other, the difference between any two corners (any side of
the simplex) is a root, and these are the only roots.  Note that
this implies six roots for $SU(3)$, since each edge of the
triangle can be traversed in two directions. So the root system of
$SU(3)$ is
  \be
\Sigma (SU(3))=  \bmu_i - \bmu_j\equiv \balpha_{ij}\ \  \
\left[{i\ne  j:\ 1 , \  2\mbox{\  or} \   3}\right]\label{sun1}
\ee

We may choose the simple roots to be $\balpha_{12}$ and $\balpha_{23}$ and
in terms of them the lattice $L_{2R}$ is defined as the set of points
 \be
 \begin{array}{l l l}
2{\bf R} &=&  2 m \balpha_{12} + 2 n \balpha_{23} \\
& \equiv & m {\bf a}_1 + n {\bf a}_2
\end{array}
\ee

We can now put the expanded root lattice together with the
minuscule representation: in Fig. \ref{Kagome} we have placed one
triangle at the origin (labeled $q$) and made copies at every
lattice point in $L_{2R}$. As the reader can see, we end up with
the kagom\'{e} lattice.

Let us note that the two features, Eqs. (\ref{sun2},\ref{sun1}),
generalize in the obvious manner for all $SU(N)$, where the
weights now point to the vertices of an $N-$simplex.

 Now we turn to some general features of our lattices valid for all the
groups we will study, not dependent on the special features of
$SU(N)$ alluded to above. To see what they might be, look at Fig.
\ref{Kagome}, and observe the following features:
    \begin{enumerate}
              \item The lattice can also be viewed as $L_{2R}$ decorated
by the conjugate (anti-quark) representation with reversed weights.
               The original representation and the conjugate share
corners, and every site is shared in this manner.
              \item If the particle can hop to any point labeled $i$ from
a point labeled $j$ in the same unit cell by moving a displacement
$\bd_{ij}$, it can keep moving an extra $\bd_{ij}$ to reach a point
labeled $j$ in the adjacent unit cell. In other words the edges of the
unit cell and the conjugate unit cell that meet at a shared corner are
continuations of each other with no change in direction.
            \end{enumerate}

            We will now furnish the proofs of these results in the general
case.

            {\bf Theorem I}:
     The original lattice $2 {\bf R}+ \bmu_i$ can be rewritten as
$2 {\bf R}+ 2 \bmu_1 -\bmu_i$.
     In other words, if at a new origin displaced from the old one by
$2\bmu_1$, we place the conjugate representation and make copies of it
using any element of $L_{2R}$, we get the old lattice. The result is just
as valid if we use any other weight $2\bmu_j$ in place of $2\bmu_1$.

    {\bf  Proof}:
     \be
    2  {\bf R} + 2 \bmu_1 -\bmu_i= 2{\bf R}+ 2 \bmu_1 -2\bmu_i + \bmu_i=
    2 {\bf R'}+\bmu_i
    \ee
    where we have used the fact that $2 \bmu_1 -2\bmu_i$, being an even
integer multiple of weight differences, is then an even integer multiple
of roots, which in turn is a translation within $L_{2R}$.

    Note that the choice of origin at $2\bmu_1$ is arbitrary: the choice
$2\bmu_2$ differs by $2\bmu_2-2\bmu_1$, an even integer multiple of roots,
and hence a translation within $L_{2R}. \ \blacksquare $

    {\bf Theorem II} If the particle can hop to any point labeled $i$ from
a point labeled $j$ in the same unit cell by moving a distance $\bd_{ij}$,
it can keep moving an extra $\bd_{ij}$ to reach a point labeled $j$ in the
adjacent unit cell.

    {\bf Proof:} Since $\bd_{ij}$ is a difference of weights it is some
integer combination of simple roots. Moving an extra distance $\bd_{ij}$,
corresponds to a total displacement by an even integer combination of
simple roots, which is a symmetry of $L_{2R}$. It follows that if we start
at a point labeled $j$ we must end up a point also labeled $j.\
\blacksquare $

\subsection{Lattices in $d=2$}

We have already discussed $SU(3)$ in the last section. Now we
will deal with $SO(5)$ and $Sp(4)$. These two Lie algebras are
mathematically equivalent up to cosmetic differences which will be
displayed.

\subsubsection{$SO(5)$} \label{SO5Lat_sec}
 We begin with the more familiar group $SO(5)$ which preserves the norm
 \be
x^2= \sum_{i=1}^{5}x_{i}^{2}
\ee
and has a defining representation  of $5\times 5$ orthogonal matrices.

We choose as Cartan generators $H_1=L_{12}$ and $H_2=L_{34}$ which
generate rotations in the $12$ and $34$ planes. In terms of the
coordinates
 \be
 x_{I}^{\pm}={x_1\pm ix_2\over \sqrt{2}}\ \ \ \
 x_{II}^{\pm}={x_3\pm ix_4\over \sqrt{2}}\ \ \
 x_0=x_5
 \ee
 we may write the invariant in this spherical basis as
 \be
 x^2= x_{0}^{2}+2\sum_{a=I}^{II}x^{-a}x_{a} \ .
 \ee

The vector $x$ itself serves as a 5-dimensional representation.
The components $x_{I}^{\pm}$ and $x_{II}^{\pm}$ are eigenstates of
$H_{1}$, $H_2$ with eigenvalues $\mathbf{H}= (\pm 1,0)$ and
$(0,\pm 1)$ while $x_0=x_5$ does not respond to either rotation
and has eigenvalues $(0,0)$. The vector representation is not
minuscule since the weights are of unequal length.

\begin{figure}[htb]
 \begin{center}
 \subfigure[]{
 \epsfxsize=1in
 \epsffile{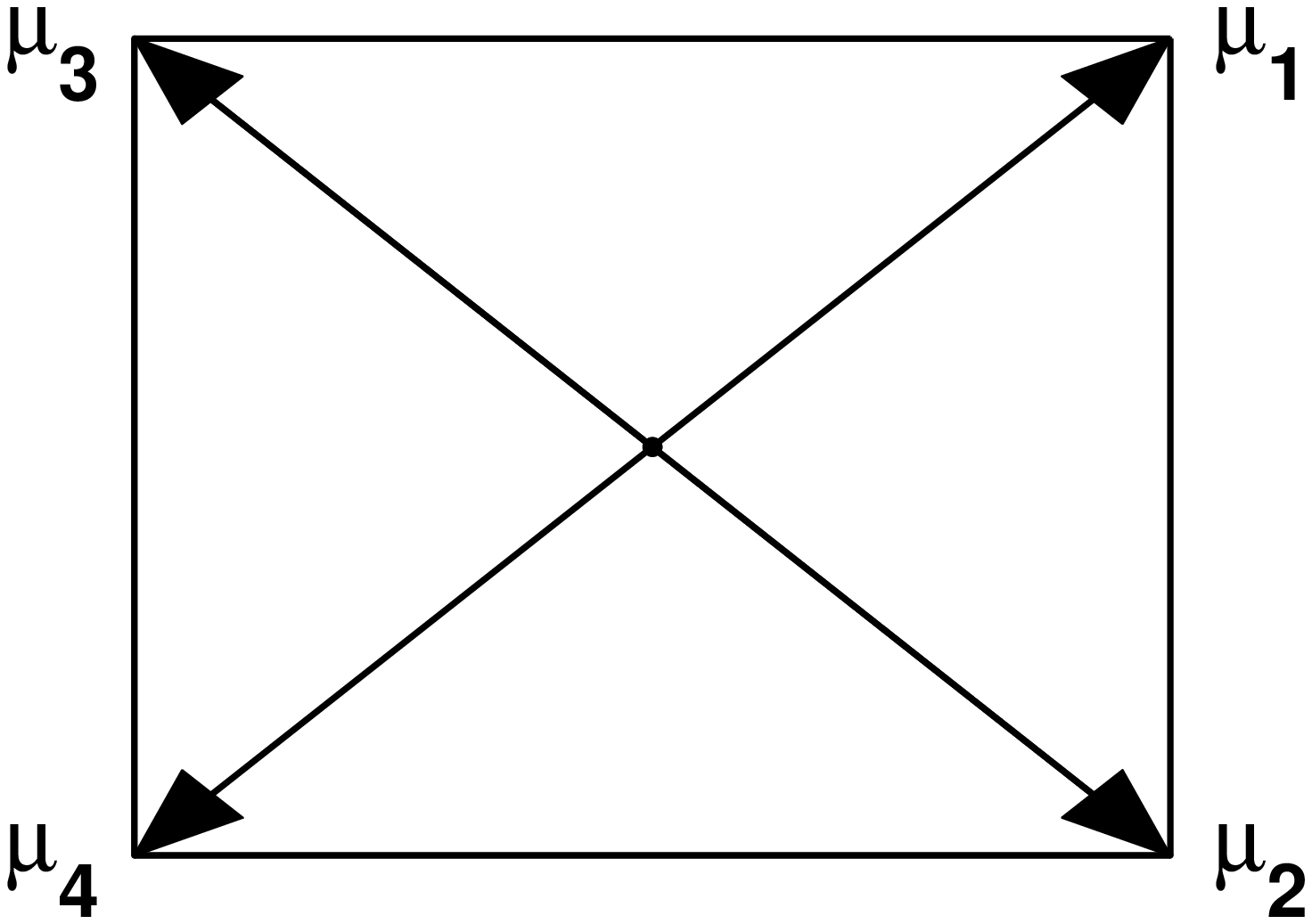}
 \label{spinorunit}}
 \subfigure[]{
 \epsfxsize=2.5in
 \epsffile{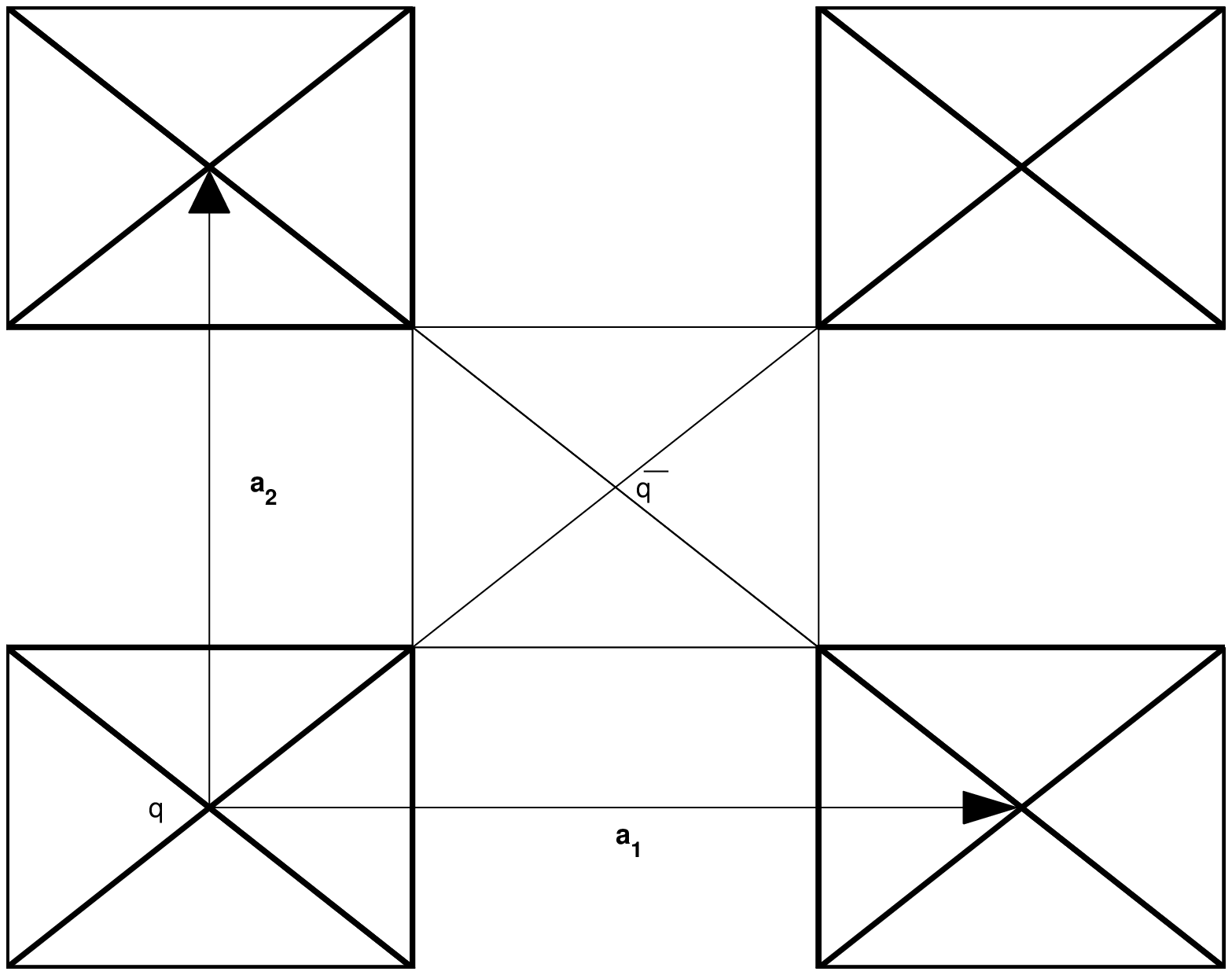}
 \label{spinorfig}}
 \caption{ Planar pyrochlore and $SO(5)$: The unit cell (a) of the spinor representation is the square of
edge unity.  The resulting square lattice, with hoppings along all
roots, is shown in (b).  The lattice vectors $\mathbf{a}_1,
\mathbf{a_2}$ are a basis for the subset of the $SO(5)$ root
lattice with even integer coefficients. }
\end{center}
\end{figure}

The only minuscule representation is the 4-component spinor, with
weights
 \be
\bmu=\left(\pm {1 \over 2}, \pm {1 \over 2}\right). \ee These form
a square as in Fig. \ref{spinorunit}.

The eight  roots for $SO(5)$ are given by
 \be
  \Sigma (SO(5))=\pm \mbe_i ; \ \pm \mbe_i \pm \mbe_j  \ \ \ \ {i\ne j=1,2}
  \ee
 where $\mbe_i$ is a unit vector in direction $i$. The short roots connect
states along the coordinate axes while the long ones go diagonally.

The same spinor also forms a representation of SO(4). However SO(4)  has
only the four long roots and one can see that the weights $\pm ({1\over
2},{1\over 2})$ do not talk to the pair $\pm ({1\over 2},-{1\over 2})$,
which means the representation is reducible. We do not discuss it here.

The simple roots are $\mbe_1-\mbe_2$ and $\mbe_2$; our root
lattice is $2 {\bf R}= 2m (\mbe_1-\mbe_2)+2n
\mbe_2=2m'\mbe_1+2n'\mbe_2$, a square lattice of sides $2$. A site
on the decorated lattice is $2 {\bf R} + (\pm {1\over 2}, \ \pm {1
\over 2})$.

Since the spinor is self-conjugate, the original squares share
corners with identical squares,  and the resulting lattice is the
square lattice.  When links along the long roots are included as
in Fig. \ref{spinorfig}, the structure is known variously as the
square lattice with crossings (SLWC), checkerboard lattice, or
planar pyrochlore. Note also that if you can hop from site $j$ to
site $i$ on one unit cell, you can hop once more by the same
amount to hit site $j$ in the next unit cell.

\subsubsection{$Sp(4)$} \label{Sp4Lat_sec}
The weights of the 4-dimensional minuscule representation of
$Sp(4)$  are \be \bmu=(\pm 1,0), ( 0, \pm 1). \ee
 Since the group is the collection of $4 \times 4$ symplectic matrices,
this is also called the defining representation. (The same
terminology applies, say to the 6-dimensional vector
representation of $SO(6)$, which is the group of $6 \times 6$
orthogonal matrices.)

\begin{figure}[htb]
\begin{center}
\subfigure{
 \epsfxsize=1in
 \epsffile{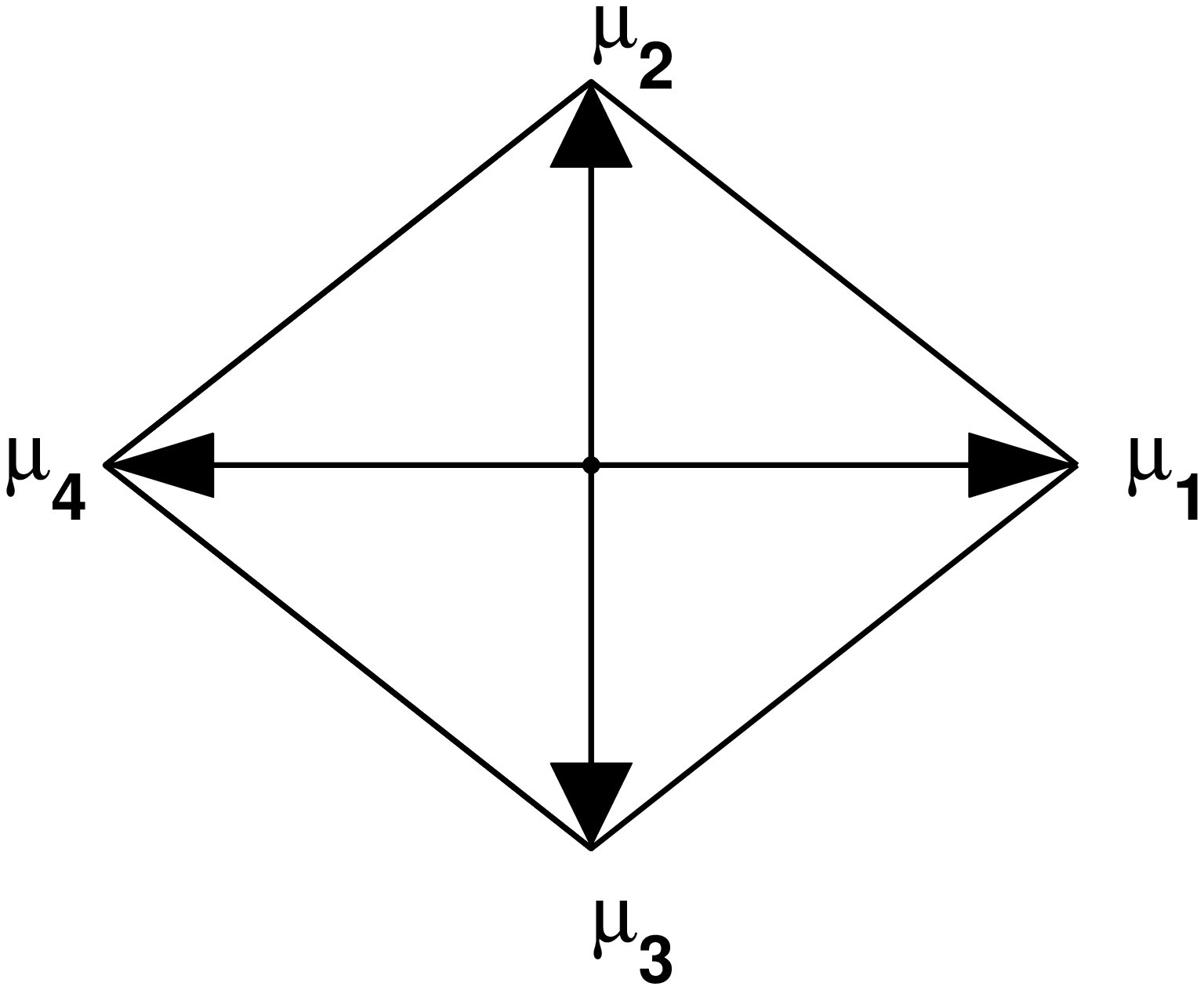}
 \label{sp4Unit}}
 \subfigure{
 \epsfxsize=2.5in
 \epsffile{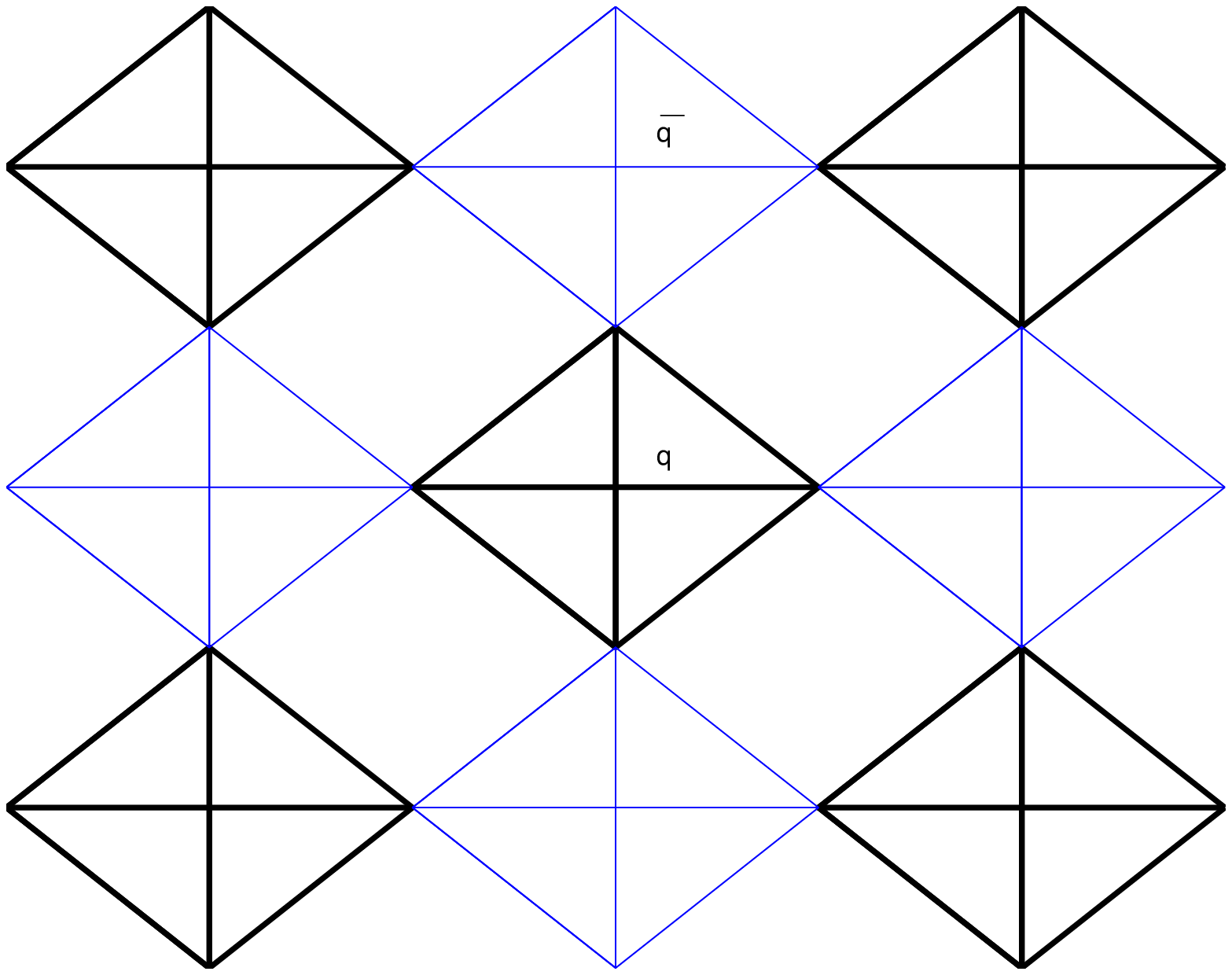}
 \label{sp4}}
 \caption{  Planar pyrochlore and $SP(4)$: The unit cell (a) for the defining representation
of $Sp(4)$ is the square rotated by 45 degrees. The entire lattice
(b) is the rotated version of the $SO(5)$ spinor. }
\end{center}
\end{figure}

The roots of $Sp(4)$ are
 \be
  \Sigma (Sp(4))= \pm \mbe_i \pm
\mbe_j\ \   \mbox{and} \pm 2 \mbe_i\ \ \   {i\ne j:\ 1 ,  2}
 \ee
 Now the long roots connect points parallel to the axes and short roots in
diagonal directions, as shown in Fig. \ref{sp4Unit}. Note that
this just the rotated and rescaled version of $SO(5)$.

This concludes the enumeration of lattices in $d=2$.

\subsection {Lattices in $d=3$}
 With the warm up from $d=2$ we can proceed rapidly through $d=3$ where
the candidates are $SU(4)$=$SO(6)$, $Sp(6)$, and $SO(7)$.
 \subsubsection{$Sp(6)$}
\begin{figure}[htb]
\begin{center}
 \epsfxsize=2.5in
 \epsffile{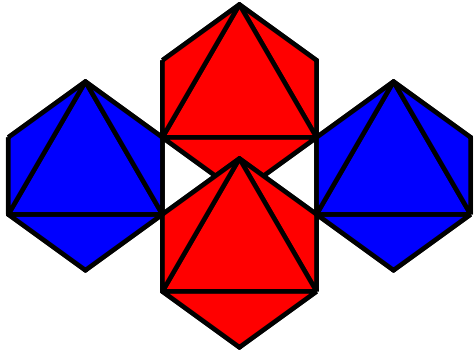}
 \caption{The octachlore lattice of $SP(6)$:  blue octagons
are copies of the defining representation of $Sp(6)$; red octagons
are in the conjugate representation.  Note that we have not drawn
the bonds along the long roots.}
 \label{sp6}
\end{center}
\end{figure}

 The only minuscule representation for $Sp(6)$ is the defining 6-dimensional
one. The weights are
 \be
 \bmu = (\pm 1, 0,0), (0,\pm 1,0), (0,0,\pm 1)
 \ee
 which form an octahedron which is self-conjugate. The decorated lattice
is most readily visualized as corner sharing octahedra, which we
will refer to as the octachlore lattice, depicted in Fig.
\ref{sp6}. The roots are
 \be
 \Sigma (Sp(6))= \pm \mbe_i \pm \mbe_j\ \ \mbox{and} \pm 2 \mbe_i\ \ \
{i\ne j:1 , 2\mbox{ or } 3}
 \ee
 The short roots allow you to hop along the edges of the octahedron while
the long roots take you straight across the unit cell to the
antipodal point.  This is just the $d=3$ version of $Sp(4)$.

 \subsubsection{$SO(7)$}
\begin{figure}[htb]
\begin{center}
 \epsfxsize=2.5in
 \epsffile{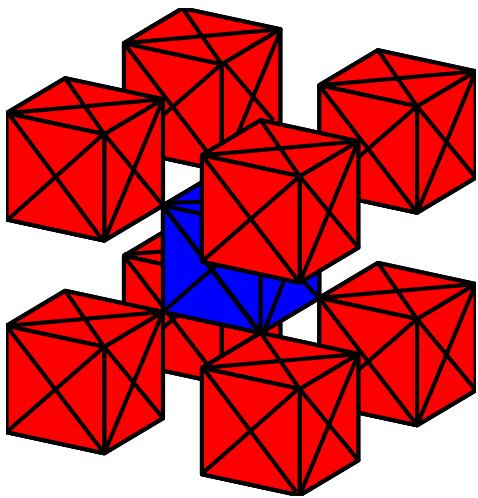}
 \caption{The 3-d checkerboard $SO(7)$ spinor lattice: blue and red cubes
show the spinor representation and its conjugate (identical in
this case).}
 \label{so7L}
 \end{center}
\end{figure}

 Next we turn to $SO(7)$. Only the spinor is minuscule. It has
self-conjugate weights
 \be
\bmu = (\pm {1 \over 2},\pm {1 \over 2},\pm {1 \over 2})
 \ee
 which lie at the corners of a unit cube. The roots
 \be
 \Sigma (SO(7))= \pm \mbe_i \pm \mbe_j\ \ \mbox{and} \pm \mbe_i\ \ \ {i\ne
j:1 , 2\ \mbox{or} \ 3}
 \ee
 allow hops along edges, and diagonally across faces, but not along the
body-diagonal. The root lattice $L_{2R}$ is cubic with edges of
size 2. The decorated lattice is made of corner sharing unit
cubes, with face diagonal hopping on every second cube as one
proceeds along any of the three cubic axes (Fig. \ref{so7L}). This
is one possible $3$ dimensional variant of the checkerboard
lattice discussed in Sections \ref{Sp4Lat_sec} and
\ref{SO5Lat_sec}; we shall refer to it as the $3$ dimensional
checkerboard lattice.  The other, which we will discuss in the
next section, is the pyrochlore.

In higher dimensions the $SO(2N+1)$ spinor leads to the
N-dimensional checkerboard lattice:  an N dimensional cubic
lattice with links on all face diagonals.

\subsubsection{$SO(6)$}
 We obtain $SO(6)$  from $SO(7)$ if we drop the short roots $\pm \mbe_i$:
\be
 \Sigma (SO(6))= \pm \mbe_i \pm \mbe_j\ \     {i\ne \ j:\ 1 , \  2\mbox{\
or} \   3}.
\ee

Consider the spinor representation of $SO(7)$. Without the short
roots $\pm \mbe_i$, we can only flip the signs of the components
of each weight two at a time. This means the 8-dimensional spinor
of $SO(7)$ breaks down into two irreducible representations with
four weights each. The first has an even number of negative
weights

 \be
\bmu_1=({1 \over 2},{1 \over 2},{1 \over 2}),\ \ \
 \bmu_2= (-{1 \over 2},  -{1 \over 2},{1 \over 2}),\ \ \
 \bmu_3= ({1\over 2},-{1 \over 2},-{1 \over 2}),\ \ \
 \bmu_4=(-{1 \over 2},{1\over 2},-{1 \over 2})
 \ee
and the other has these weights reversed and hence an odd number
of negative weights.  This is general: the irreducible spinor of
$SO(2N+1)$ becomes two irreducible representations of $SO(2N)$,
called left and right handed spinors.

\begin{figure}[htb]
\begin{center}
 \epsfxsize=2.5in
 \epsffile{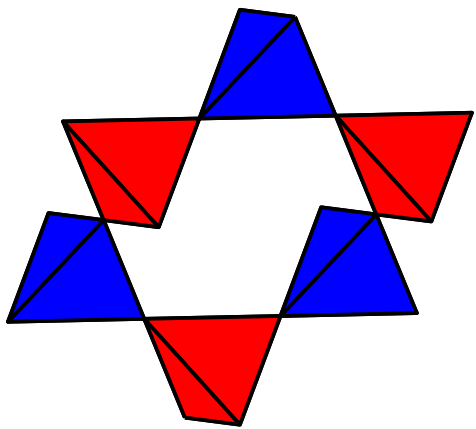}
 \caption{The pyrochlore $SO(6)$ spinor lattice:   Blue tetrahedra are
in the fundamental representation of $SU(4)$, or the right-handed
spinor representation of $SO(6)$; red tetrahedra are in the
conjugate representation.}
 \label{PyroLat}
\end{center}
\end{figure}

If you join the 4 points in either spinor multiplet, you will see
the tetrahedra that form the weights of the $SU(4)$ quark and
antiquark representations. In other words the right and left
handed spinors of $SO(6)$ are the quark and anti-quark of $SU(4)$.
(This is why we will not study $SU(4)$ separately.) The tetrahedra
form the familiar pyrochlore lattice shown in Fig. \ref{PyroLat}.

The extensions of this construction to higher dimensions yield
corner sharing simplices that form the natural generalizations of
the kagom\'{e} and pyrochlore lattices \cite{TorquatoStillinger}.

  Since for $SU(4)$ the difference of any two weights of the quark is a root
and there are no others, the roots of $SO(6)$ may just as well be
written as
  \be \label{PyroBasis}
\Sigma (SO(6))=  \bmu_i - \bmu_j\ \  \    \left[{i\ne  j:\ 1 , \
2, \ 3, \mbox{\  or} \   4}\right]
 \ee
a result we will invoke later.

  The third minuscule representation of $SO(6)$ is the 6-dimensional
defining representation with weights
  \be
\bmu = (\pm 1, 0,0),\ \ \ (0,\pm 1,0), \ \ \ (0,0,\pm 1) \ee
 just as in $Sp(6)$. The decorated lattice again is made of corner-sharing
octahedra, as shown in Fig. \ref{sp6}. However, without the long
roots $\pm 2\mbe_i$, one can hop only along the edges but cannot
jump directly from a point to its antipodal point.

This concludes the enumeration of lattices.
 \subsubsection{Structure factor}
 Before we turn on the flux let us note that the structure factors for
these lattices have a simple group theoretic interpretation. When we do a
sum over all sites (indexed by $l$) we find
 \be
 \begin{array}{l l l l l}
S(\bk )&= & \sum_{l\in  L_{2R}+\bmu}e^{i \bk \cdot \br_l} \ \ \
           & = \sum_{L_{2R}}e^{i \bk \cdot 2\bR} \sum_{j\in \bmu}e^{i \bk\cdot
           \bmu_j} \ \ \
           & = \sum_{L_{2R}}e^{i \bk \cdot 2\bR} {\rm Tr} e^{i \bk \cdot\mathbf{H}} \\
 &\equiv & S_{L_{2R}}(\bk ) \chi(\bk )
 \end{array}
 \ee
 where {$ \mathbf{H}_i = H_i$, the $i^{th}$ element of the
Cartan subalgebra, whose $j^{th}$ eigenvalue is $ \bmu_j^i$,} and
$\chi(\bk ) = {\rm Tr} e^ {i \bk \cdot \bH} $ is just the character of the
representation.

\section{ Flux Hamiltonians in $d=2$.} \label{2dFluxSection}
 We will turn on fluxes by attaching arrows to each bond. The sense of the
arrow will remain fixed as we move along bonds in any one
direction. The hopping amplitude will be $\pm i$ if we go along
(against) the arrow. Time reversal reverses every arrow, sending
$H \to -H$. Since time reversal also reverses $\bk$, this means
that in cases we study,
 \be
H(-\bk ) = -H(\bk ).
\ee

If all hopping amplitudes are pure imaginary, the flux through
each triangle is $\pm \pi /2$. It is important to remember that
the arrows themselves do not stand for physical quantities. For
example a tetrahedron with uniform flux $\pi /2$ coming out of
each face is invariant under all symmetries of the tetrahedron
although the arrows will look different if we say rotate the
figure. This will, of course, be built into a PSG (Projective
Symmetry Group) analysis \cite{WenPSG}.

\subsection{$SU(3)$ and kagom\'{e}} \label{KagFlux_sec}

\begin{figure}[htb]
\begin{center}
\epsfxsize=2.5in
\epsffile{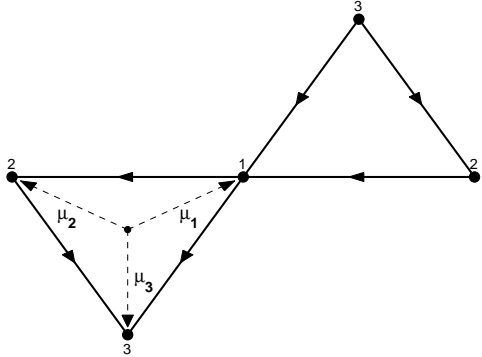}
 \caption{Flux assignment on the kagom\'{e} lattice. There is a phase factor of $\pm
i$ as we move along (against) the arrow.}
 \label{kagomeflux}
\end{center}
\end{figure}

Let us begin with the first case, the kagom\'{e} lattice, now with
the flux shown in Fig.\ref{kagomeflux}. Note that as we go
counter-clockwise around the (quark) triangles 1-2-3-1, we get a
product $(i)(+i)(-i)=+i$. This is so for every quark triangle. The
antiquark triangles will have the opposite flux.

Consider hops from the site numbered $1$ in the central unit cell
shown, to the sites numbered $2$ to its left (on the same cell)
and right (on the cell to the right). Hopping along (against) the
arrows brings a factor of $\pm i$. We note that this state was one
of several mean-field solutions found by \cite{MarstonZeng} for
uniform hopping on the kagom\'{e} lattice.

In terms of creation and destruction operators $c$ and $c^{\dag}$ we have,
in obvious notation, a contribution to $H$ :
 \be
 \begin{array}{l l l}
H_{1,2} (\bk )&=& c_{2}^{\dag}c_1 \left (i e^{i\bk \cdot (\bmu_2-\bmu_1)}+ h.c \right )  - (1 \leftrightarrow 2)\\
&=& (c_{2}^{\dag}c_1+c_{1}^{\dag}c_2 ) \sin (\bk \cdot
\balpha_{12})
\end{array}
 \ee
 where we have dropped an overall factor of $2$, suppressed the $\bf k$
dependence of the operators, and as before,
 \be
\balpha_{12}=\bmu_1 -\bmu_2
\ee
 Upon noting that $c_{1}^{\dag}c_2=E_{\balpha_{12}}$, the generator
corresponding to the root $\balpha_{12}$, we see that when all hops are
included we get what we shall the {\em canonical} form
 \be
H(\bk ) =\sum_{\Sigma^+} \sin (\bk \cdot \alpha)\left[ E_{\balpha}
+ E_{\balpha}^{\dag}\right]
 \ee
where the sum is over $\Sigma^+$, the positive roots, which have
the form $\mu_i - \mu_j$ with $i < j$.

In this paper we will often modify the canonical form in two ways: replace
$\sin( \bk \cdot \alpha)$ by $\bk \cdot \alpha$ and attach various signs
$s_{\balpha}=\pm 1$ in front of each term so $H$ assumes the more general
form
 \be \label{GeneralH}
  H(\bk , s_{\balpha})=\sum_{\Sigma^+} s_{\balpha} \sin (\bk \cdot \balpha ) (E_{\balpha}+E_{\balpha}^{\dag}).
 \ee
The role of these signs is to modify the relative phase of the
hopping amplitudes on various bonds, and therefore the fluxes.

 Putting in explicit values, we get for the canonical case
 \be H(\bk )= \begin{bmatrix}
0 &\sin x&\sin (\half (x+ \sqrt{3} y)) \\
\sin x & 0& -\sin (\half (x- \sqrt{3} y))\\
\sin (\half (x+ \sqrt{3} y))&-\sin (\half (x- \sqrt{3} y))&0
\end{bmatrix}
 \ee
 {\em where $x$ and $y$ stand for $k_x$ and $k_y$ respectively.}

 The determinant of this matrix
 \be
 |H|= -2 \sin x \sin \left(\frac{1}{2} \left(x-\sqrt{3}
   y\right)\right) \sin \left(\frac{1}{2}
   \left(x+\sqrt{3} y\right)\right)
   \ee
 shows zeros along the lines $x=0$ and $x=\pm \sqrt{3}y$, which are
precisely the directions of the weights (or their negatives)! The
determinant also has a simple form in terms of the three positive roots:

 \be
  |H |= -2 \sin (\bk \cdot \balpha_{12}) \sin (\bk \cdot \balpha_{13}) \sin (\bk \cdot \balpha_{23})
 \ee

 We do not know how this generalizes for $SU(N)$. But we do know how to
understand the lines of zeros as follows.

 The Hamiltonian has the form
 \be
 \sum_{i<j}^{} \sin (\bk \cdot (\bmu_i -\bmu_j))(E_{\balpha_{ij}}+E_{\balpha_{ij}}^{\dag})
 \ee
 Suppose we set $\bk =\bmu_1$. (All points on the simplex are the same and
we pick one that is easier to analyze.) The simplex has the property that
 \be
 \bmu_{i} \cdot \bmu_j =
\begin{cases}
a & i=j \\
 b  & i \ne j
\end{cases}
 \ee

 It follows that the argument of any sine in which $\bmu_1$ does not
appear will vanish and the ones where it does will have the same value
$a-b$. The resulting matrix, proportional to $\sin (a-b)$,  has
non-zero entries only in the first row or column.  This leaves a $ 2
\times 2$ submatrix of zeros which kills the determinant:
 \be
 \sin ( a-b)  \left|\sum_{j=2}^{3}  (E_{\balpha_{1j}}+E_{\balpha_{1j}}^{\dag})\right|=0
 \ee

 This will happen for all $SU(N)$ because each raising or lowering operator
in the fundamental representation has only one non-zero entry.
(Geometrically, this is equivalent to the statement that the N-simplex has
no parallel edges).  The components of the Hamiltonian which do not vanish
for $\bk = \bmu_1$ are
 \be
\sin (a-b) \sum_{j=2}^{N} (E_{1j}+E_{1j}^{\dag}) \ee
 But all of these are hoppings to or from site 1, so that all non-zero
entries lie in either the first row or the first column of the
matrix. The resultant $(N-1)\times (N-1)$ null submatrix ensures
that the determinant is zero at this particular value of $\bk$.

Let us now replace $\sin x$ by $x$, since none of the key features are
lost and the algebra is more manageable, especially in the problem of
diagonalization. So we will set

\be
H(\bk )=
\left(
\begin{array}{lll}
 0 &  x & \frac{1}{2} \left(x+\sqrt{3} y\right)  \\
x & 0 &
   -\frac{1}{2} \left(x-\sqrt{3} y\right) \\
  \frac{1}{2} \left(x+\sqrt{3} y\right) & -\frac{1}{2} \left(x-\sqrt{3} y\right) & 0
\end{array}
\right)
\ee

 The determinant of this matrix is
 \be
 | H|= -\half x (x-\sqrt{3}y)(x+\sqrt{3}y)
 \ee
 which shows zeros along the same lines $x=0$ and $x=\pm \sqrt{3}y$ as
before.

 If we change the sign of the term multiplying any of the generators, we
get the same lines of zeros. This is expected since we simply
reverse the flux penetrating every triangle, which inverts the
spectrum. If we reverse the signs of any two of them, it makes no
difference even to the flux. The reader can check the lines of
zeros are not altered by any change of sign.

 As explained earlier, this is a problem where $H(\bk )=-H(-\bk )$, and
the pair of points at $\pm \bk$ together produce $E\to -E$
symmetry. With the Fermi energy at zero we are at half-filling,
the relevant filling for mean-field solutions of the Heisenberg
model.

The energies themselves are fairly complicated and not displayed
here. It turns out two of them never vanish away from the origin
and one of them produces all the zeros: if we move around the unit
circle, it vanishes six times. Linearizing near these zeros will
produce one-dimensional Dirac fermions that will control the
low-energy physics.  ( Near the origin all six Dirac excitations
will get mixed up.) If all this were part of a mean-field
calculation, we would be looking at this Dirac field minimally
coupled to a gauge field if we wanted to consider fluctuations.
One could ask if the mean-field solution remains stable in their
presence. These questions will be considered separately
\cite{wip}.

 \subsection{Flux Hamiltonians for $Sp(4)$=$SO(5)$}
 \label{SO5Flux_sec}
 Recall that the groups $Sp(4)$ and $SO(5)$ are the same. The roots of one are
the rotated and rescaled versions of the other and no new physics
will come from looking at both.  We will only work with $SO(5)$
since it may be more familiar to the reader.

 Before writing down the hopping matrix we need to define the basis. The
states are numbered $1$ through $4$ in Fig. \ref{so5} with $1=
(\half ,\half )$ etc. We use a tensor product of two Pauli
matrices, $\sigma$ and $\tau$, to operate on the two labels. We
take as generators
 \be
 \begin{array}{l l l}
 E_{\mbe_1}&=& \half \sigma_+\otimes I_{}= E_{-\mbe_1}^{\dag}\\
 E_{\mbe_2}&=& \half \sigma_3\otimes \tau_+=E_{-\mbe_2}^{\dag}\\
 E_{\mbe_1 \pm \mbe_2}&=& \mp \left[E_{\mbe_1},E_{\pm \mbe_2}\right]=\pm \half \sigma_+\otimes \tau_{\pm}\\
 E_{-\mbe_1 \mp \mbe_2}&=&E_{\mbe_1 \pm \mbe_2}^{\dag}
 \end{array}
 \ee

\subsubsection{The canonical Hamiltonian}
\begin{figure}[htb]
\begin{center}
\epsfxsize=1.75in
\epsffile{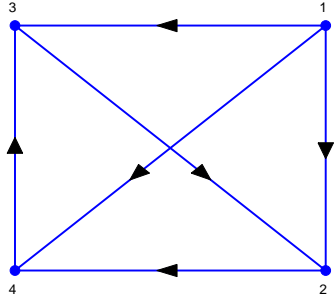}
 \caption[The flux on the $SO(5)$  lattice]{Flux assignment for the canonical $SO(5)$ Hamiltonian.
There is a phase factor of $\pm i$ as we move along (against) the
arrow. The flux in each triangle alternates as we go
counterclockwise in the case depicted.}
 \label{so5}
\end{center}
\end{figure}

 In this basis we choose the canonical form
 \be
 \begin{array}{l l l}
 H(\bk ) &=& \sum_{\Sigma_+} (\bk \cdot \balpha)(E_{\balpha } + E_{\balpha }^{\dag})\\
 &=& \left(
\begin{array}{llll}
 0 & y & x & x+y \\
 y & 0 & y-x \ \ & x \\
 x & y-x \ \ & 0 & -y \\
 x+y \ \ & x & -y & 0
\end{array}
\right) \ . \end{array} \label{so51}
 \ee
The orientation of the arrows corresponding to this $H$ are shown
in Fig. \ref{so5}. In other words, rather than write down some
arrows and deduce the hopping matrix from these, we are writing
down a canonical matrix in the Lie algebra and asking what hopping
elements it implies.

 If we look at the flux in each triangle we find that it alternates:
triangles sharing a face diagonal, such as triangles $132$ and
$342$ in Fig. \ref{so5}, have the same flux, while triangles not
sharing a face diagonal (triangles $132$ and $142$) have opposite
flux. The determinant here is
 \be
|H|=4 (x^4 - x^2 y^2 + y^4)
\ee
 It has permutation symmetry but not full rotational symmetry. It has no
zeros anywhere away from the origin. We do not know a simple way
based on group theory to understand this.

 The characteristic polynomial is
 \be
 E^4 - 4 E^2 (x^2 + y^2) + 4 (x^4 - x^2 y^2 + y^4)
 \ee resulting in the particle-hole symmetric spectrum:
\be
E = \pm \sqrt{2} \sqrt{ x^2+y^2 \pm \sqrt{3} x y }
\ee
 The reason behind the{ symmetry $ E\rightarrow -E$} is  the matrix
 \be
 G= \left(
\begin{array}{llll}
 0 & 0 & 0 & \ \ 1 \\
 0 & 0 & \ 1 \ & \  0 \\
 0 & -1 \ \ & 0 & \  0 \\
 -1 \ \ & 0 & 0 & \ 0
\end{array}
\right) \ ,
 \ee
with $G^2 = -I$.  Since
 \be
 G\cdot H \cdot G^{-1}=-H
 \ee
 it follows that $H$ and $-H$ have the same spectrum.

  The anatomy of the operator $G$ is interesting. Suppose we wanted to
manufacture an operator that reversed the sign of $H$ by
conjugation.  We could accomplish this by a parity operation that
exchanges each weight with its negative- this should flip every
hopping term. However there are many ways to flip the weights
since we can take each state to its parity reversed state, times
any unimodular phase factor, which must be a sign if we want the
hoppings amplitude to be $\pm i$. Suppose we picked a $G'$ with
all positive signs:
  \be
 G'=
\left(
\begin{array}{llll}
 0 & 0 & 0 &\   1 \\
 0 & 0 &\  1 \ & 0 \\
 0 & \ 1 \ & 0 & 0 \\
 \ 1 \ & 0 & 0 &  0
\end{array}
\right)
\ee

   We would find
  \be
  G' H G'^{-1}= \left(
\begin{array}{llll}
 0 & -y & x & x+y \\
 -y & 0 & y-x \ \ & x \\
 x & y-x \ \ & 0 & y \\
 x+y  \ \ & x & y & 0
\end{array}
\right)
\ee
 This is clearly not $-H$. However, if we go back to the lattice and ask
what the corresponding hopping amplitudes are we will find that
all the fluxes are reversed, though some arrows are reversed and
some are not. Since $-H$ also has all fluxes reversed (reversing
every bond will reverse the product over every triangle) the two
must be gauge equivalent. It turns out that appending minus signs
to states $3$ and $4$ is one way to flip the arrows that needed to
be flipped.  The operator $G$ is the product of $G'$ and a
diagonal matrix that multiplies $3$ and $4$ by minus signs. This
is to be expected in a gauge theory, where the symmetry is
projectively realized \cite{WenPSG}.

  Since such a procedure will work for any self-conjugate representation,
we will not explicitly construct the operator in future occasions.

  We can understand now why  the spectrum of $H$ had the full
set of lattice symmetries even though the flux alternated. Under
any of the lattice symmetry operations, we either left the flux
alone or reversed it. Neither affects the determinant since these
operations at worst exchange $E \to -E$, which has no effect on
the spectrum. This feature will be seen again when we consider
other groups.

 Since $H$ is $4\times 4$, and the characteristic equation is even in $H$,
it satisfies an equation of the form
 \be \label{GenD}
\left ( H^2 - f(k) \cdot I\right) ^2 = g(k) \cdot I
\ee
 where $f$ and $g$ are scalar functions. Eq. (\ref{GenD}) reduces to
the Dirac form if $g(k) =0$ and $f$ is constant.

\subsubsection{The non-canonical Hamiltonian}

Let us now change the sign in front of any of the terms in Eq.
\ref{so51}. It turns out that the determinant is sensitive only
the relative sign of the two long roots that reach diagonally
across the square. Here is what we get when we flip the
coefficient of the $E$ corresponding to the root $\mbe_1-\mbe_2$:
 \be
H_{uni}(\bk )= \left(
\begin{array}{llll}
 0 & y & x & x+y \\
 y & 0 & -(y-x) & x \\
 x & -(y-x) & 0 & -y \\
 x+y & x & -y & 0
\end{array}
\right) \label{so52}
\ee

If we compute the flux now, we find it is uniform in all triangles, hence
the subscript in $H_{uni}$. This in turn means that $H_{uni}$ will be
invariant (up to gauge transformations) under symmetry operations of the
lattice.

The characteristic polynomial is
\be
E^4 - 4 E^2 x^2 - 4 E^2 y^2 + 4 x^2 y^2
\ee
 resulting in the spectrum
\be
 E = \pm \sqrt{2} \sqrt{x^2+y^2 \pm \sqrt{x^4+x^2 y^2+ y^4} }
\ee
 which has $E\to -E$ symmetry because $G$ once again anticommutes
with $H$. The spectrum has zeros along the lines $x=0$ and $y=0$.
Note that along these directions, $\balpha_{ij} \cdot \mathbf{e}_x
= 1$ or $0$.  As in the kagom\'{e} case, the lines of zeros are
axes of symmetry of the unit cell. All edges not orthogonal to
these axes have equal projections onto them (up to sign).  With
signs as in $H_{uni}$, this results in a pair of identical rows in
the Hamiltonian -- and thus lines of zeros along the short roots
for both the linearized {\em and} lattice versions of $H$.

\subsubsection{The  Hamiltonian with unequal hopping}

In the above we have considered a more general case
\be
  H(\bk , s_{\balpha})=\sum_{\Sigma^+} s_{\balpha} \sin (\bk \cdot \balpha ) (E_{\balpha}+E_{\balpha}^{\dag}) \\
 \ee
 where $s_{\balpha}=\pm 1$ is a possible sign. In this problem there were
essentially just two choices, the ones with uniform and alternating
fluxes, determined by the relative sign of the two long hops.

 In a problem like $SO(5)=Sp(4)$, where there are roots (i.e. bonds) of two
different lengths, we could also play with the relative strengths
of the hopping across long and short bonds. There is no obvious
inspiration from group theory on how to choose from the continuum
of possibilities,  though only some choices will give $SU(N)$
mean-field solutions. The only consolation is that only two
different lengths are allowed for the roots of any semi-simple Lie
algebra and among the cases we study this happens only for
$SO(2N+1)$ and $Sp(2N)$.

 We just mention one extreme case where long hops are set equal to zero:

 \be
H_{short}(\bk )= \left(
\begin{array}{llll}
 0 & \ y \ & \ x \ & 0 \\
 \ y \ & 0 & 0 & \ x \ \\
 \ x \ & 0 & 0 & \ -y\  \\
 0 &\  x \  & \ -y \  & 0
\end{array}
\right) \label{so53}
\ee

Now the determinant is
 \be
 |H_{short}|= (x^2 + y^2)^2
 \ee
 which describes two Dirac points at the origin.   Indeed, this is
just the flux phase on the square lattice originally described by
\cite{AffleckMarston}. The unit cell is of course twice as big as it needs
to be, so that the two Dirac points of the traditional unit cell have both
come to the origin.

More generally, with a magnitude $c$ for the long hops and
relative signs all positive, the eigenvalues are
 \be
  E = \pm \sqrt{(1+c^2)(x^2+y^2) \pm 2 c x y \sqrt{c^2+2} }
 \ee
and the spectrum has one Fermi point at the origin. For relative
signs chosen as in $H_{uni}$, the eigenvalues are
 \be
 E= \pm \sqrt{(1+c^2) (x^2+y^2) \pm 2 c \sqrt{ x^4+y^4 +c^2 x^2 y^2} }
 \ee
 At the special values $c=0, \sqrt{2}$, these give a Dirac spectrum.
 For all $c \neq 1$, the Fermi surface is a single point at the
origin.  Values of $c$ corresponding to mean-field solutions are
given in Section \ref{MFSection}.

\section{Flux Hamiltonians in $d=3$} \label{3dFluxSection}

 Luckily we have to consider just three groups: $SO(6)$=$SU(4)$,
$SP(6)$, and $SO(7)$. The only minuscule representation of the
latter is the spinor. There are three minuscule representations
for $SO(6)$: two spinors (quark and antiquark of $SU(4)$) and the
six-dimensional vector representation.  $SP(6)$ has one minuscule
representation-- the defining one.

\subsection{Flux on the $SO(6)$ spinor lattice.}

 The weights forming the tetrahedron are
 \be \label{PyroWeights}
\bmu_1 = ({1 \over 2},{1 \over 2},{1 \over
2}), \ \ \
 \bmu_2 = (-{1 \over 2},-{1 \over 2},{1 \over 2}),   \ \ \
\bmu_3 = ({1 \over 2},-{1 \over 2},-{1 \over 2}), \ \ \
 \bmu_4 = (-{1 \over 2},{1 \over 2},-{1 \over 2})
 \ee
The positive roots are, in terms of orthogonal unit vectors,
 \be
\mbe_1\pm \mbe_j\ \ \ \ \ \ \ j>i=1,\ 2,\ 3.
 \ee

However, as pointed out in Eq. (\ref{PyroBasis}), it is more
convenient to note that since this is also an $SU(4)$ quark
representation, we could write them in terms of the weights
(\ref{PyroWeights}) as

\be
\balpha_{ij}^{+}=\bmu_i -\bmu_j\ \ \ \ \ j>i.
\ee
 In view of what we saw in $d=2$ we are going to admit the more general
case

\be
  H(\bk , s_{\balpha})=\sum_{\Sigma^+} s_{\balpha} \sin (\bk \cdot \balpha ) (E_{\balpha}+E_{\balpha}^{\dag})\label{signs}
 \ee
 where $s_{\balpha}=\pm 1$ is a possible sign in front of each term.

\subsubsection{The canonical Hamiltonian}

\begin{figure}[htb]
\begin{center}
\subfigure[]{
 \epsfxsize=2.5in
 \epsffile{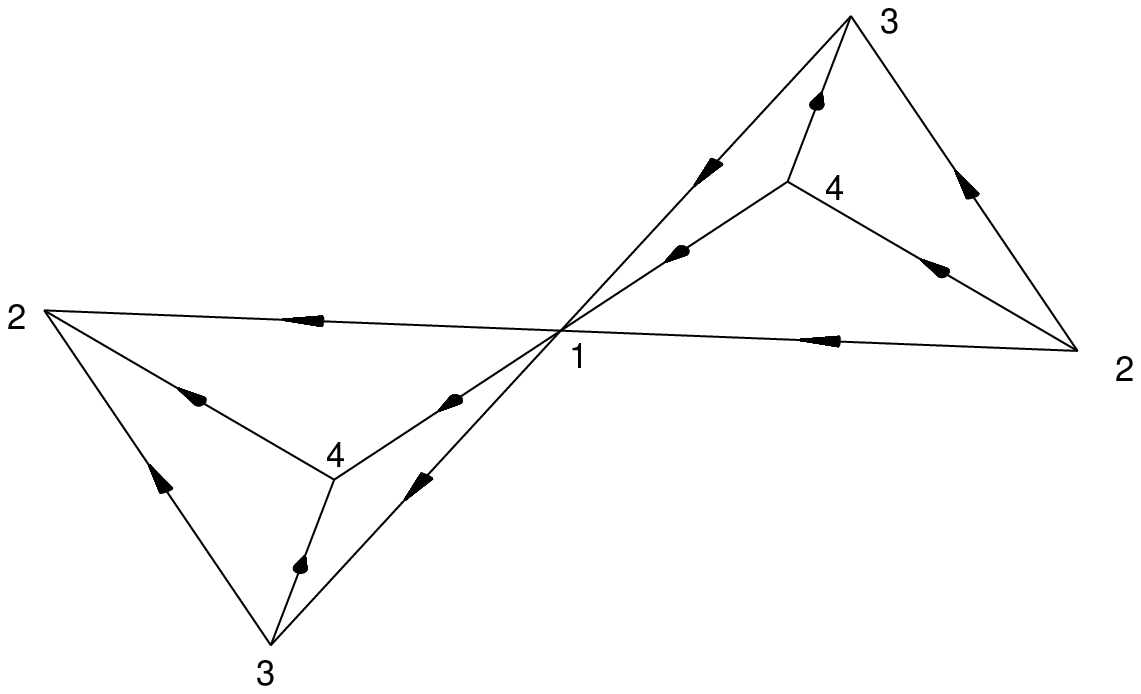}
 \label{monofig2}}
 \subfigure[]{
 \epsfxsize=2.5in
 \epsffile{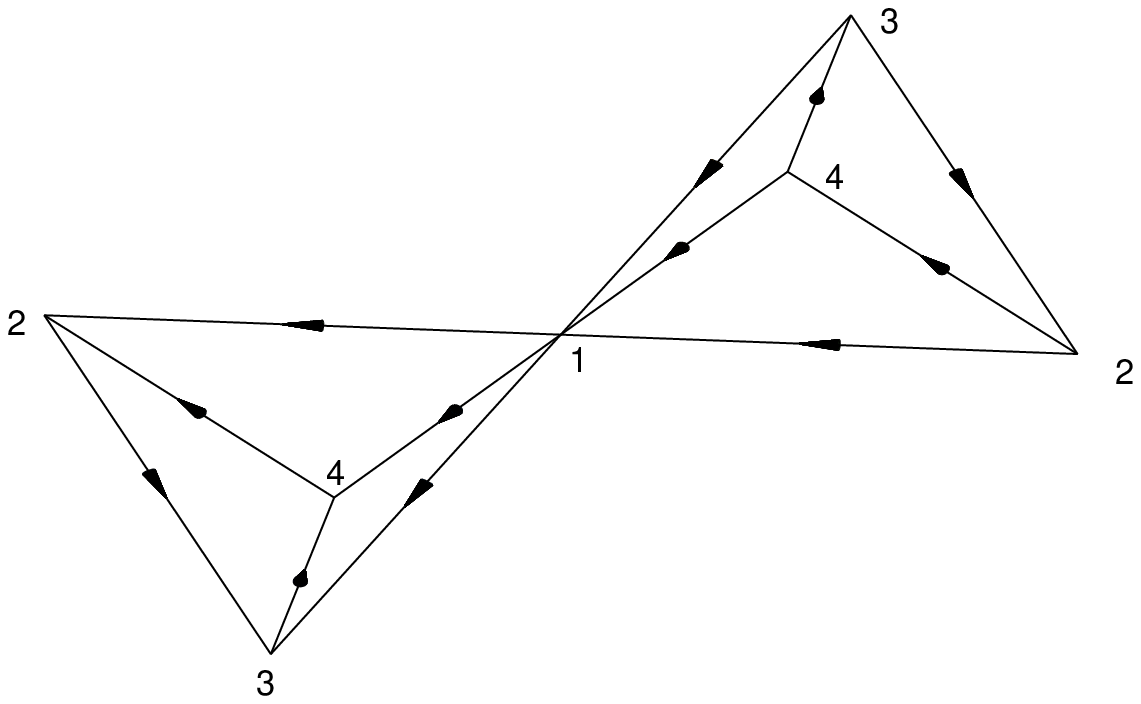}
 \label{monopolefig}}
 \caption{Possible flux
assignments to the pyrochlore.  (a) Flux assignment breaking
rotational symmetries. (b) Flux assignment preserving rotational
symmetries.}
\end{center}
\end{figure}

 If we pick all signs positive, (which means the arrow always goes from a
site with a lower index to one with a higher index) we obtain the
flux assignment in Fig. \ref{monofig2}.  This gives the canonical
Hamiltonian
 \be
 H=\left(
\begin{array}{llll}
 0 & x+y \ & y+z\  & x+z \\
 x+y \ & 0 & z-x \ & z-y \\
 y+z \ & z-x \ & 0 & x-y \\
 x+z \ & z-y \ & x-y \ & 0
\end{array}
\right)
\ee
with determinant
\be
|H|=4 x^4 - 4 x^2 y^2 - 4 x^2 z^2 + 4 y^2 z^2
\ee
 Note that it lacks the discrete symmetries of the lattice. This is to be
expected since the flux on each face is not the same.

Consider its zeros. It vanishes along the weight directions, $\bk
\propto \bmu_i$. This is to be expected since the right-handed
spinor is also the $SU(4)$ quark representation and we have seen
that for $SU(N)$, because the weights form a simplex, when $\bk
\propto \bmu_i$ only terms corresponding to roots involving
$\bmu_i$ remain (and that all have the same coefficient in front).
The rest vanish, so that $H$ has just one nonzero row or column.

But we find in addition that there are entire planes along which there are
zeros. For example for any linear combination $\bk = a \bmu_1+b\bmu_2$ or
$\bk = a \bmu_1+b\bmu_4$, the determinant vanishes. However it does not
vanish for $\bk = a \bmu_1+b\bmu_3$ unless $a=0$ or $b=0$. This variation
is to be expected since the flux is not symmetric on the tetrahedron.

Once again if we can use more powerful group theoretic methods to
know when determinants of certain elements of the Lie algebra of
the type of Eq. (\ref{signs}) will vanish, we will be able to
anticipate this result rather than just observe it.

\subsubsection{The uniform monopole case} \label{MonFlux_sec}
If we flip the
 sign of the $24$ and $42$ matrix
elements (corresponding to the root $ \bmu_2-\bmu_4=\balpha_{24}$)
we obtain the flux assignment of Fig. \ref{monopolefig}. The
Hamiltonian is

 \be
 H_{mono}=\left(
\begin{array}{llll}
 0 & x+y \ & y+z \ & x+z \\
 x+y \ & 0 & z-x \ & -(z-y) \\
 y+z \ & z-x \ & 0 & x-y \ \\
 x+z \ & -(z-y)\  & x-y \ & 0
\end{array}\label{h24}
\right)
\ee
with determinant
\be
|H_{mono}|=4 x^4 +  4 y^4 +4z^4 - 4 x^2 y^2 - 4 x^2 z^2 - 4 y^2 z^2
\ee
 which has the discrete symmetries of the lattice. The subscript on $H_{mono}$ reflects the fact that the
flux is the same on all faces of the tetrahedron and comes from  a unit monopole at its center.

The reader may well ask how many more such signs are we going to play
with. Luckily we are done.

To understand this, we need to transcribe the Hamiltonian to the
corresponding factors of $\pm i$ on the edges of the tetrahedron. As
mentioned above, the case with all $s_{\balpha} =1$ has a factor of $i$ if we
move from a corner to another with larger index (and a $-i$ if we move the
other way). It is readily verified that the two faces not involving the
bond $24$ have an outward flux of $\half \pi$ (a factor $i$ around the
triangular faces) and the other two the reverse. Clearly the choice of
matrix elements violates the discrete symmetries of the tetrahedron.

On the other hand if we flip the coefficient of the $\balpha_{24}$ term, we get an arrangement with all
outward fluxes equal to $\pi /2$ and we are led to $H_{mono}$, with a
tetrahedrally symmetric determinant.

Other choices of sign will only yield one of two options: the flux is
uniform (could be $\pm \half \pi$) and the determinant is symmetric, or
the flux assignment breaks the symmetry with two positive and two negative
faces. Different choices for the latter will correspond to determinants in
which the asymmetric roles of $x$, $y$ and $z$ are interchanged.

We will now elaborate further on the case $H_{mono}$, which
describes uniform flux, as it has various nice properties. It has
been discussed elsewhere as an interesting mean-field Hamiltonian
for the $SU(2)$ Heisenberg model on the pyrochlore lattice
\cite{shoibal-thesis,bcs-toappear}. To make contact with existing
literature on this problem, we will briefly revert to the custom
of referring to momentum components as $k_x$ or $k_y$  rather than
simply $x$ or $y$.

The energy levels of $H_{mono}$  are
\begin{equation} \label{MonoE}
E(k) =\pm\sqrt{2\sum_i
k_i^2\pm 2\sqrt{3\sum_{(i<
j)}k_i^2k_j^2}}
\end{equation}
The spectrum has $E \rightarrow -E$ symmetry since there is a
matrix $G$ obeying $G^2=I$ that anticommutes with $H_{mono}$:
 \be
H_{mono} G= -G  H_{mono}
\ee
where
\be
G =\frac{1}{\sqrt{3}}\begin{bmatrix}
 0&\ 1\ &\ 1\ & \ 1 \ \\
\ -1 \ &   0 & \ 1\ & \ -1 \ \\
\ -1 \ &\  -1\ &  0 & \  1 \ \\
\ -1\ & \ 1 \ & \ -1 \ &  0 \\
 \end{bmatrix}
 \ee
  That a matrix $G$ which ensures $E\to -E$ symmetry  should occur is less obvious than in the $SO(5)$ case,
since  tetrahedron
is not inversion symmetric (self-conjugate).  The inversion operation maps the
tetrahedron formed by the right-handed spinor representation of
$SO(6)$ to that formed by the left-handed representation.
Hence $G$ is not a simple geometric operation on sites in the unit
cell.

We are tempted to   cast $H_{mono}$ in Dirac form \be
H_{mono}=\a_xk_x+\a_yk_y+\a_zk_z \ee since $G$ seems to be like
the matrix  $\beta$  which anticommutes with the three $\a$'s in
the Dirac Hamiltonian. However the resemblance to the Dirac case
is not complete because  $\a$'s do not form a Clifford algebra
and $H^2_{mono}$ is  not a multiple of the unit matrix.

What one finds is
 \be  \label{acom1}
 \begin{array}{l l l}
\left[ \a_i , \a_j \right]_+&=& 2\delta_{ij} +  \sqrt{3}|\varepsilon_{ijk}|W_k  \\
\left[ W_i , W_j \right]_+&=& 2\delta_{ij}.
 \end{array}
 \ee
In other words, the anticommutator of the $\a$'s is  proportional
to the unit matrix plus some amount of $W$'s, and the $W$'s obey a
Pauli algebra. Thus if we square $H (\a )$, move the stuff
proportional to the unit matrix  to the left hand side and square
again, we will end up with a multiple of the unit matrix. Indeed
this is so:
 \be \label{bcom}
(H^2 -2 k^2 ) ^2 =12 ( k_x^2 k_y^2 + k_x^2k_z^2 + k_y^2 k_z^2)
 \ee
where the subscript on $H$ and the identity $I$ have been
suppressed.

That we should end up with the form  encountered in the $d=2$
$SO(5)$ case of Eq. (\ref{GenD})
 \be
(H^2-f(k)\cdot I )^2=g(k)I\label{cast}
 \ee
is due to  the same reasons:  the characteristic  polynomial
$P(H)$ is even and of fourth order in $H$,  i.e.  quadratic in
$H^2$.  It can therefore be cast in the form Eq. (\ref{cast}). To
get all details of $f$ and $g$ we would of course need to actually
evaluate $P(H)$:
 \be
P(H)= H^4 - 4 H^2 (k_{x}^2 + k_{y}^2 + k_{z}^2) + 4 (k_{x}^4 +
k_{y}^4  + k_{z}^4 - k_{x}^2 k_{y}^2 - k_{y}^2 k_{z}^2 -k_{z}^2
k_{x}^2))=0.
 \ee

The anticommutator algebra in  Eqs. (\ref{acom1}) stems from the
fact that  the Hamiltonian lives in the Lie algebra of the right
handed spinors of  $SO(6)$, which is also the quark of SU(4).

Recall that  it is possible to write the generators of SO(N) in
the spinor case in terms of the Dirac $\gamma$- matrices:
$\sigma_{\mu \nu}$, which generates rotations in the $\mu-\nu$
plane, may be expressed as $\sigma_{\mu \nu}={i\over
2}\gamma_{\mu}\gamma_{\nu}$. Although the $\gamma$ matrices are $8
\times 8$, bilinears in them like $\sigma_{\mu \nu}$ form
reducible representations with two  $4\times 4$ blocks, these
being the quark and antiquark of SU(4). The two blocks are
eigenstates of $\gamma_7=i \gamma_1 \cdots  \gamma_6$ with
eigenvalue  $\pm 1$.  If we want the quark we can work with these
$8 \times 8$ matrices and focus on just the top left hand corner.
In this block $\gamma_7$ is just a number equal to $1$.

Consider the following operator
 \be
H= i\gamma_1(\gamma_6-\gamma_4)k_x+
i\gamma_3(\gamma_2+\gamma_6)k_y +i\gamma_5
(\gamma_4+\gamma_2)k_z=H_R \oplus H_L
 \ee
where $H_R$ and $H_L$ are $4 \times 4$ blocks corresponding to
right and left handed spinors, or quark and antiquark
representations.

With a judicious choice of basis for the $\gamma$ matrices its
upper left-hand corner, $H_R$ is just our Hamiltonian
$\a_xk_x+\a_yk_y+\a_zk_z$. Thus if we do not stray from this block
we can view the $\a$'s  as bilinears of $\gamma$ matrices. Not so
obvious is the fact that the $W$'s which come from two powers of
$\a$ are also bilinears in $\gamma$.

The closure under anticommutation of the $\sigma_{\mu \nu}$ or the
$\a$'s and $W$'s is a  special property of SO(6). In general, if
you multiply two of them you will get something quartic in the
$\gamma$'s even after  some of them reduce to quadratic terms upon
invoking $\gamma^2=I$. The quartic ones can be rewritten as
$\gamma_7$ times a quadratic, upon inserting the square of the
``missing" two $\gamma$ matrices. In the sector with $\gamma_7=1$,
these are just quadratic in the $\gamma$'s.

\subsection{Flux on the $SO(6)$ vector lattice.}
  In the defining vector representation the generators are represented as
follows in terms of canonical creation and destruction operators $c$ and
$c^{\dag}$:
 \be
 \begin{array}{l l l}
 H_i&=& c^{\dag}_{i}c_i- c^{\dag}_{-i}c_{-i}\ \ \  i=1, 2, 3\\
 E_{\mbe_i\mp \mbe_j}&=& c^{\dag}_{i}c^{}_{\pm j}- c^{\dag}_{\mp j}c_{-i} \ \ \ i<j\le 3\\
 \end{array}
 \ee
 with generators of negative roots defined as the adjoints of the positive
ones above.

\subsubsection{The canonical Hamiltonian} \label{SO6Vec_sec}

\begin{figure}[htb]
\begin{center}
\epsfxsize=2.5in
\epsffile{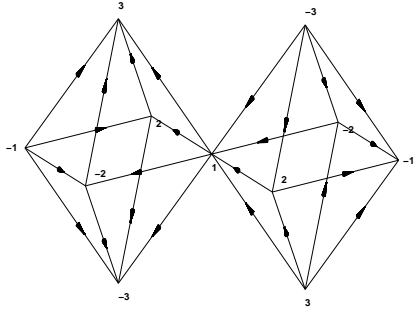}
 \caption{Flux assignment to the octachlore in accordance with signs of
the group generators of the vector representation of $SO(6)$.}
 \label{hexafig}
\end{center}
\end{figure}

 In this basis the usual sum over positive roots with all coefficients
positive yields the matrix
 \be \label{so6vecHam}
 H=\left(
\begin{array}{llllll}
 0 & 0 & x-y & x+y & x-z & x+z \\
 0 & 0 & -x-y & y-x & -x-z & z-x \\
 x-y & -x-y & 0 & 0 & y-z & y+z \\
 x+y & y-x & 0 & 0 & -y-z & z-y \\
 x-z & -x-z & y-z & -y-z & 0 & 0 \\
 x+z & z-x & y+z & z-y & 0 & 0
\end{array}
\right)
\ee
 where the rows and columns are numbered as follows: $(1,0,0,)\equiv 1,
(-1,0,0)\equiv -1, (0,1,0)\equiv 2,...(0,0,-1)\equiv -3$, the
components being just the eigenvalues of $H_1,\ H_2 $ and $H_3$.
The site labels and corresponding factors of $\pm i$ are shown in
Fig. \ref{hexafig}.  Note that the flux alternates from one face
to the next.

The determinant vanishes identically because there are two zeros at every
$\bk$. If we pull them out we find
 \be
|H|=48 ( x^2 y^2 + y^2 z^2 + z^2x^2 )*0*0
\ee
 that is to say, the product of the nonzero energies is $ 48 ( x^2 y^2 +
y^2 z^2 + z^2x^2 )$.

Why does $H$ have all the discrete symmetries when the flux
alternates? The answer is that any rotation is equivalent to a
change of the sign of the overall flux, which in turn corresponds
to time-reversal, and does not affect the determinant in a problem
with $E \to -E$ symmetry.

   Extra zero-energy bands occur when any two coordinates vanish,
i.e., along the axes, which corresponds to the direction of the
weights. We can understand this to the extent we could understand
the $SU(N)$  and $SO(5)$ cases. If
 \be
H= \sum_{i<j} (\bk \cdot (\mbe_i \pm \mbe_j)) \left[ E_{\mbe_i \pm
\mbe_j}+E_{\mbe_i \pm \mbe_j}^{\dag}\right]
 \ee
it follows that if we set $\bk = \mbe_1$ say, only roots of the
form $\mbe_1 \pm \mbe_j$ will survive and that too with the same
coefficient. The matrix will have only two non-zero rows and
columns -- for the sites at $ \pm \mbe_1$ in the unit cell.  With
the flux assignment of (\ref{so6vecHam}), one row is exactly the
negative of the other, resulting in two extra zero energy bands in
both the linearized and the lattice Hamiltonian.

 Near any line of zeros we can define a $2$ dimensional Dirac field,
except near the origin when they all collide and modify each other.

Again there is a matrix $G$ which anticommutes with the
Hamiltonian, and acts upon the unit cell as the inversion.  Its
existence results from the fact that the unit cell is inversion
symmetric, while the directions of all fluxes are reversed by
inversion.

\subsubsection{The non-canonical Hamiltonians}

We could append signs for each term,  but this gives spectra which
break the lattice symmetries. We have not looked deeply into what kind of
zeros result in that case.

\begin{figure}[htb]
\begin{center}
\epsfxsize=2.5in
\epsffile{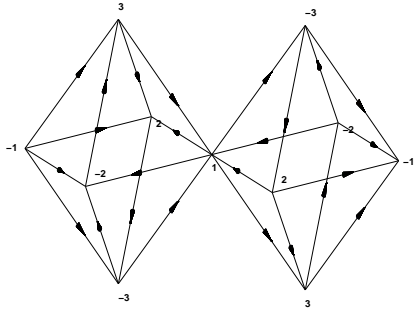}
 \caption{Flux assignment to the octachlore preserving rotational
symmetries.}
 \label{hexafigsym}
\end{center}
\end{figure}

We did however note the following. Suppose we start with a hopping
problem on an octahedron with uniform flux in every face as in
Fig. \ref{hexafigsym}.

When we extracted the $H(\bk )$ for that problem we found it could
not be written in terms of generators of $SO(6)$. It is important
to understand why we have this problem here but did not when we
considered the spinor of $SO(6)$ = quark of $SU(4)$. There each
root or generator connected only two states. If we did not like
the sign of the matrix element given by group theory we just put a
negative sign in front of that generator using $s_{\balpha}$. But
here, each root connects two pairs of points, corresponding to
parallel edges of the octahedron. For example
 \be
E_{\mbe_1+\mbe_2}= c^{\dag}_{1}c_{-2}- c^{\dag}_{2}c_{-1}
 \ee
connects points labeled $(-2, 1)$ and $(-1,2)$ in Fig.
\ref{hexafigsym} with {\em opposite} matrix elements. If we do not
like the relative sign, we cannot do anything about it.  This is
exactly what happens in the case of the octahedron with uniform
flux in every face. To describe it, we would have to use the
generators of the much larger group SU(6). But that is not the
game we are playing: we want to work within a group, $SO(6)$ being
the operative one here.

This could have happened to the $SO(5)$ spinor, whose short roots
connected opposite sides of the square with same sign for the
horizontal roots $\pm \mbe_1$ and opposite signs for the vertical
roots $\pm \mbe_2$. Luckily this choice of signs corresponded to
the case of interest.

\subsection{$Sp(6)$ }
 The defining representation of $Sp(6)$ is the same octahedron as in $SO(6)$
with the same weights. The generators can be written in terms of creation
and destruction operators as

\be
 \begin{array}{l l l}
 H_i&=& c^{\dag}_{i}c_i- c^{\dag}_{-i}c_{-i}\ \ \  i=1, 2, 3\\
 E_{\mbe_i\mp \mbe_j}&=& c^{\dag}_{i}c^{}_{\pm j}\mp  c^{\dag}_{\mp j}c_{-i} \ \ \ i<j\le 3\\
 E_{2\mbe_i} &=& 2 c^{\dag}_{i}c_{-i}
 \end{array}
 \ee
 with negative roots being given by adjoints of the above.

\subsubsection{The non-canonical Hamiltonian} \label{SP6Flux2_sec}
 It is interesting to consider first the canonical $H$ with all signs
positive and {\em only the short roots.} (As noted before when we have two different root lengths, we have the freedom to chose the scale of each type of term. Keeping only short roots is an extreme case.) We find
 \be
 H_{short}(\bk )=
\left(
\begin{array}{llllll}
 0 & 0 & x-y & x+y & x-z & x+z \\
 0 & 0 & x+y & y-x & x+z & z-x \\
 x-y & x+y & 0 & 0 & y-z & y+z \\
 x+y & y-x & 0 & 0 & y+z & z-y \\
 x-z & x+z & y-z & y+z & 0 & 0 \\
 x+z & z-x & y+z & z-y & 0 & 0
\end{array}
\right)
\ee

The determinant has the value
\be
|H_{short}|=-32 (x^2 + y^2) (x^2 + z^2) (y^2 + z^2)
 \ee
 that is to say, zeros along the weights. The logic is the same as in
$SO(6)$ since the long roots that distinguish between them have
been suppressed. Note however that matrix elements are different
now: the two pairs of states connected by a generator do not
always have opposite matrix elements.   Thus in this case along
the axes there are 2, rather than 4, zero energy bands.

What is surprising is that the energies do not change {\em for any
choice of signs !}
 \subsubsection{A non-canonical Hamiltonian with unequal
coefficients} \label{SP6Flux1_sec}

 Consider the following matrix involving the long
roots:

\be \label{HSP6_2}
 H_{ ``x+z",\half} \left(
\begin{array}{llllll}
 0 & 2 x & x-y & x+y & x-z & -x-z \\
 2 x & 0 & x+y & y-x & -x-z & z-x \\
 x-y & x+y & 0 & 2 y & y-z & y+z \\
 x+y & y-x & 2 y & 0 & y+z & z-y \\
 x-z & -x-z & y-z & y+z & 0 & 2 z \\
 -x-z & z-x & y+z & z-y & 2 z & 0
\end{array}
\right)
\ee
 where the subscripts remind us of two ways in which it differs from the
canonical form: the $x+z$ term has a minus sign relative to the
canonical form, and the hopping matrix element for the long roots
is half as big as the canonical one. As for the latter point,
consider the term $2x$. It indeed equals $\bk \cdot 2\mbe_1$, but
the generator $E_{2\mbe_1}= 2c^{\dag}_{1}c_{-1}$ has another two
in it. So this term should have been $4x$. But with the choice of
sign and hopping (\ref{HSP6_2}) we get
 \be
|H_{ ``x+z",\half}|= -16 (x + y)^2 (x + z)^2 (y + z)^2
 \ee
which has zeros along planes $x+y=0$ etc.

When we put in the canonical strength ($4x$ etc.) we did not find
any interesting spectra for many choices of sign that we tried.

\subsection{$SO(7)$ spinor}

Recall that the only minuscule representation of $SO(7)$ is the
spinor and that the lattice we associate with it is cubic, with
face diagonals but no body diagonals (Fig. \ref{SO7fig1}). The
generators in this representation can be expressed in the direct
product space of three Pauli matrices:

 \be
 \begin{array}{l l l l l l}
 E_{\pm \mbe_1} &=& \frac{1}{2}\sigma_3 \otimes \tau_3 \otimes \alpha_\pm  \ \ \ \
 & E_{\pm \mbe_2}& = & \frac{1}{2}\sigma_3 \otimes \tau_\pm \otimes 1   \\
  E_{\pm \mbe_3} &=&  \frac{1}{2}\sigma_\pm \otimes 1 \otimes 1 \ \ \ \
 &E_{\mbe_1\pm \mbe_2} &=& \pm\frac{1}{2}1 \otimes \tau_+ \otimes \alpha_{\pm}  \\
 E_{\mbe_1\pm \mbe_3} &=& \pm  \frac{1}{2}\sigma_+ \otimes \tau_3 \otimes \alpha_{\pm} \ \ \ \
 & E_{\mbe_2\pm \mbe_3} &=&  \pm \frac{1}{2}\sigma_+ \otimes \tau_{\pm} \otimes 1
 \end{array}
 \ee
 where the labels $1, 2$, and $3$ correspond to the directions of the co-
ordinate axes.

The matrix
\be
G = \sigma_2 \otimes \tau_1\otimes \alpha_2
 \ee
 acts as an inversion operator on the unit cell and anti-commutes with
{\em all} of the symmetric generators $E_{\a} + E_{-\a}$ so that the
spectrum has symmetry under $E \rightarrow -E$ no matter what the signs.

\subsubsection{The canonical Hamiltonian}

 If we ask what hopping amplitudes are associated with the canonical case
we find that each square plaquette has $\pi$ flux.

The generators $E_{i \pm j}$ fix the flux through the triangular
plaquettes to be $\pm \pi /2$.  The form of these generators dictates that
two opposing pairs of triangular faces will have diagonals with the same
orientation; the third will have diagonals with opposite orientations.
It turns out that in this case a uniform flux through the triangular
plaquettes is impossible.

If we choose all signs to be positive, then three of the cube's faces have
flux $\pi/2$ outwards through all triangular plaquettes, and the remaining
three to have flux $- \pi/2$.

The zeros of energy can be found from
\be
|H| = \frac{1}{256} (x^2+y^2+z^2-2(x z-y z -x y) )^2(x-y+z)^4
\ee

Here not all cubic symmetries are preserved, but permutations of the $x$,
$-y$, and $z$ axes (corresponding to rotations of the cube about the $(1,
-1, 1)$ body diagonal) map positive fluxes to positive fluxes and vice
versa, so that some of the cubic symmetries are preserved.

\subsubsection{The non-canonical Hamiltonian with alternating
flux}  \label{SO7Alt_sec}

\begin{figure}[htb]
\begin{center}
\subfigure[]{
 \epsfxsize=2in
 \epsffile{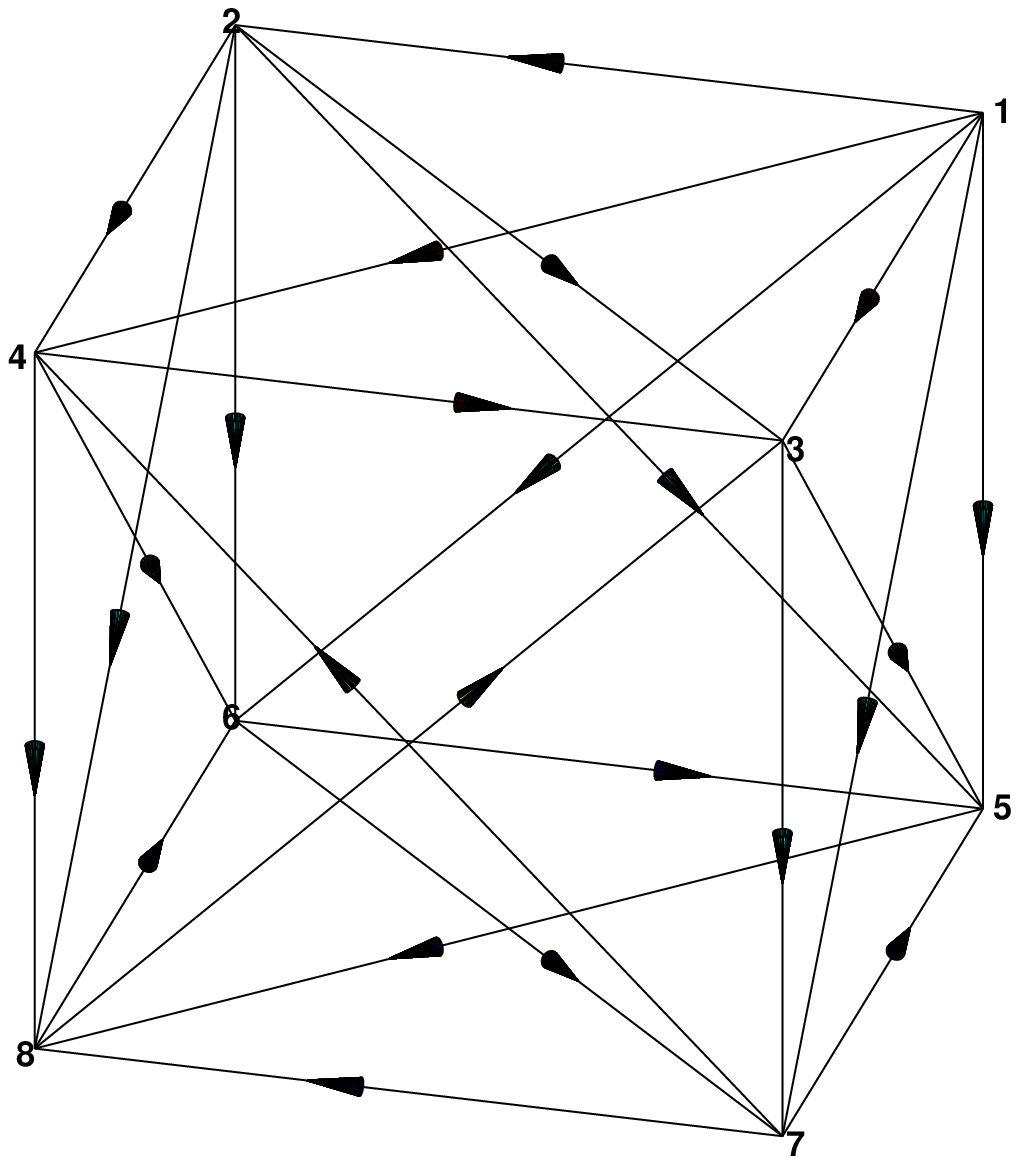}
 \label{SO7fig1}}
 \subfigure[]{
 \epsfxsize=2in
 \epsffile{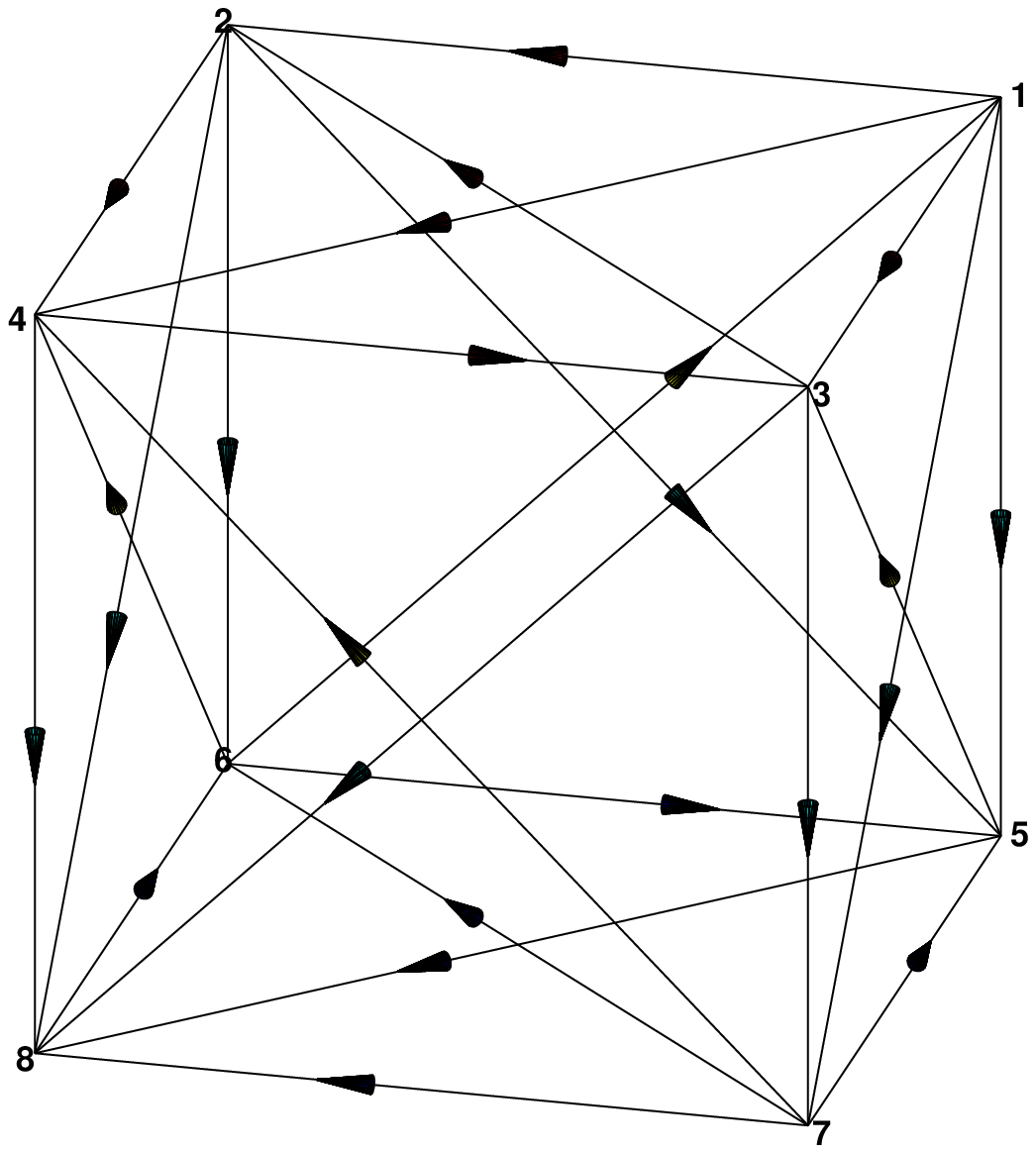}
 \label{SO7fig2}}
\end{center}
 \caption{Two possible flux assignments for on the $SO(7)$ lattice. Each
square plaquette has flux $\pi$, while the triangular plaquettes
on each face may be chosen to have equal (a) or alternating (b)
flux.}
 \end{figure}

As in the octahedral case, the other interesting case is the
alternating flux pattern shown in Fig. \ref{SO7fig2}, in which
rotations by $\pi/4$ about the $x$, $y$, and $z$ axes reverse the
signs of all fluxes. In this case, we append minus signs to the
$x+z$, $x-y$, and $y-z$ terms. The energies are remarkably simple:

\be
E= \pm \frac{1}{2} \sqrt{3} ( x \pm y \pm z )
 \ee
which vanish along the planes $ x = \pm y \pm z$.  For example, if
the sites of the unit cell are labeled $1 -8$ as shown in Fig.
\ref{SO7fig2}, momenta in the plane $ x = -y-z$ obey $ \bk \cdot
\balpha_{1j} = - \bk \cdot \balpha_{8j}$, producing a pair of
linearly dependent rows in the Hamiltonian. In the alternating
flux case the cubic symmetries are preserved since a $\pi /4 $
rotation reverses the signs of all fluxes, which is gauge
equivalent to reversing the signs of all hoppings and thus the
sign of $H$. Invariance under $E \rightarrow -E $ thus ensures
that this is a symmetry.

These two possibilities are the only ones preserving the permutation
symmetry of the $x$, $y$ and $z$ axes.

\section{Comments} \label{CommentsSection}

\subsection{Relevance to mean-field solutions of the Heisenberg model} \label{MFSection}

We will now revisit the Hamiltonians of Sections
\ref{2dFluxSection} and \ref{3dFluxSection} with a view to asking
whether they are, in fact, mean-field solutions of the $SU(N)$
Heisenberg model.  A few of these have been discussed previously
in the literature.  The kagom\'{e} Hamiltonian of Section
\ref{KagFlux_sec} was identified as a mean-field solution by
\cite{MarstonZeng}.   Several authors have discussed $SU(N)$
mean-field states on the square lattice with second neighbour
hopping \cite{WWZ,LaughlinZou,HermeleTc,LaughlinZou2}, though
these have focused on the gapped chiral spin state quite unlike
that of Section \ref{SO5Flux_sec}; we believe that previous
mean-field studies on the checkerboard lattice \cite{Bernier} have
been restricted to dimerized states.

We wish to extend these results and establish that many of the
Hamiltonians discussed above are mean-field solutions. To show
this, we must argue that the mean-field equations admit solutions
in which the hoppings $t_{\alpha}$ are purely imaginary, and
hoppings along roots of equal length are equal in magnitude.

The mean-field equation can be written \cite{LaughlinZou}
 \be  \label{MFCrit}
t_{ij} = \frac{J_{ij}}{\pi^2}\int_{q \in 1BZ} \sum_{E^{(n)}(\bk) <
0} \frac{1}{|E^{(n)}(\bk)|} H_{ij}(\bk) e^{-i \bk \cdot \br_{ij}}
 \ee
where $J_{ij}$ is the spin-spin coupling between sites $i$ and $j$
for the original Heisenberg Hamiltonian, and $t_{ij}$ is the
hopping matrix element between these sites for the mean-field
Hamiltonian. If the spectrum is invariant under $k \rightarrow
-k$, then $H(k) = - H(-k)$ guarantees that the real part of the
integral vanishes. Further, if the spectrum preserves the lattice
symmetries, then the integral will clearly have the same magnitude
for all roots $\alpha_{ij}$ of a given length.  In cases with two
different length roots, the mean-field equations specify a
particular relative hopping strength.

The reader should be warned, however, that some of the
Hamiltonians we have studied have large manifolds of zero energy,
such as planes of zeros. These may lead to a divergent integral in
(\ref{MFCrit}), in which case the Hamiltonian is {\em not} a
mean-field solution.

Such exceptions aside, many of the flux configurations discussed
in Sections \ref{2dFluxSection} and \ref{3dFluxSection} do clearly
give self-consistent mean-field Hamiltonians. Consider first the
lattices in which all links have the same length (and are related
by lattice symmetries): the kagom\'{e}, pyrochlore, and
octachlore. In these cases (\ref{MFCrit}) gives equal hopping
amplitudes on all links provided that the spectrum does not break
the lattice symmetries. The kagom\'{e} Hamiltonian has been
discussed by \cite{MarstonZeng}; it is somewhat exceptional among
the Hamiltonians we consider in that its spectrum is not inversion
symmetric at fixed $\bk$.  The monopole Hamiltonian of the
pyrochlore lattice (Section \ref{MonFlux_sec}), as well as $SO(6)$
(Section \ref{SO6Vec_sec}) and short-root $SP(6)$ (Section
\ref{SP6Flux2_sec}) Hamiltonians on the octachlore lattice have
spectra that preserve lattice symmetries and thus naively should
be mean-field states by the argument above.  We have verified
numerically that the monopole and short-root $SP(6)$ Hamiltonians
are viable mean-field solutions.  The $SO(6)$ Hamiltonian on the
octachlore lattice, however, has two bands of zero energy, causing
(\ref{MFCrit}) to diverge; it is not a mean-field solution.

The lattices constructed from the spinor representation of
$SO(2N+1)$, as well as the canonical lattice of $SP(2N)$, have
roots of two different lengths. In all three examples discussed
here, Hamiltonians with hopping only along the short roots are
mean-field solutions; for $SO(5)$ this gives the flux state of
\cite{AffleckMarston} and for $SO(7)$ the 3-d version thereof.  If
we are interested in Hamiltonians with non-zero hopping along the
long roots, we must find the ratio $t_l / t_s$ consistent with Eq.
(\ref{MFCrit}).

A summary of the allowed values of $t_l /t_s$ is given in Table
\ref{MFTab}. For $SP(6)$ generic values of $t_l/t_s$ do not give
symmetric spectra, and we find no mean-field Hamiltonians with
$t_l >0$.  The alternating flux Hamiltonian of $SO(7)$ has a
symmetric spectrum for general $t_l / t_s$; however for many $t_l
/t_s$ two-dimensional surfaces of zero energy cause the integral
to diverge and we are unable to find a mean-field solution with
$t_l>0$.  Both flux assignments discussed in Section
\ref{SO5Flux_sec} for the $SO(5)$ spinor lattice give
symmetry-preserving spectra whose mean-field $t_l /t_s$ can be
calculated numerically.  The ratio $t_l/t_s$ at mean-field depends
on the relative magnitudes of the spin-spin coupling $J_l/J_s$ in
the original Heisenberg Hamiltonian; solutions with $t_l >0$ exist
only for sufficiently large $J_l$, as shown in Table \ref{MFTab}.

\begin{table}[ht]
\begin{center}
 \begin{tabular}{|c|c|c|c|c|}
\hline
& \ \  $SO(5)$ (alt)\ \ & \ \ $SO(5)$ (uni) \ \ & \ \ $SO(7)$ (alt)\ \ &\ \  $SP(6)$  \ \ \\
\hline $J_l/J_s$ & $t_{l}/t_{s} $&  $t_{l}/t_{s} $& $t_{l}/t_{s} $ & $t_{l}/t_{s} $ \\
$1$ &  $0, 1.59 $  &   $0,  2.28  $ & $0$ & $0$\\
$.9$ & $ .371 $ & $1.66 $ &$0$  & $0$\\
$ .8$ &  $ 0 $ & $1.28 $ & $0$ & $0$\\
$ .7 $ & $ 0 $ & $1.07 $ & $0$ & $0$ \\
 \hline
\end{tabular}
\label{MFTab}
\end{center}
 \caption{ Relative strengths of hopping along the long ($t_l$) and short
($t_s$) roots as determined by the mean-field equations for the
$SO(5)$, $SO(7)$, and $SP(6)$ hopping problems.  For $SP(6)$ and
$SO(7)$ we find only the $t_l =0$ solution at mean-field level.
For $SO(5)$ we find consistent mean-field solutions with $t_l>0$
for $J_l/J_s > .682$ in the uniform flux case, and $.891$ in the
alternating flux case. }
 \end{table}

Needless to say, this analysis does not preclude the existence of
other flux assignments leading to a lower mean-field energy.
Indeed states with lines and planes of zeros, as many of our
examples have, are often energetically disfavoured at mean-field
\cite{MarstonZeng} due to the large phase space near $E=0$
relative to gapped or mostly gapped states. Also one must bear in
mind that dimerized mean-field states of lower energy inevitably
exist \cite{Dimers,ReadSachdev}.  However, as pointed out in
Section \ref{LargeNIntro}, corrections to the mean-field solution
for $N< \infty$ often alter the relative stability of various
mean-field states, so we should not take this issue too seriously.
Among the mean-field solutions discussed here, an interesting
example of this is the monopole Hamiltonian of the pyrochlore
lattice.  It corresponds to the lowest energy symmetry preserving
mean-field solution to the $SU(N)$ Heisenberg model, and has lower
energy than the dimerized mean-field ground states after
Gutzwiller projection is used to enforce the constraint of single
occupancy \cite{bcs-toappear}.

\subsection{Hamiltonians beyond the linear approximation}

This work has focused on the linearized versions of lattice
Hamiltonians, in which we have replaced
 \be \label{LatSub}
\sin ( \bk \cdot \br ) \rightarrow \bk \cdot \br  .
 \ee
However, many of the interesting properties of the spectra are
unaffected by this substitution.

First, surfaces of zero energy which are related to symmetries of
the unit cell will not be affected.  Recall that we find several
zero-energy surfaces along directions of the unit cell for which
$\bk \cdot r_{ij}$ takes on values $\pm c, 0$ for some constant
$c$. If $\bk \cdot r_{ij}$ is replaced by $ \sin(\bk \cdot
r_{ij})$, the only effect on the Hamiltonian at these points is to
change the value of the constant $c$; hence the zero eigenvalues
remain.  As discussed above, this yields lines of nodes along the
weight vectors for the $SU(N)$ lattice model for any $N$.  The
zero-energy manifolds of the alternating flux $S0(5)$ and $SO(7)$
Hamiltonians, and the $SO(6)$ and $SP(6)$ Hamiltonians discussed
above for the octachlore lattice, are also preserved under
(\ref{LatSub}). In other words, all zero-energy surfaces listed in
Sections \ref{2dFluxSection} and \ref{3dFluxSection} which reflect
symmetries of the lattice unit cell are unaffected by the
substitution (\ref{LatSub}).

Second, the symmetry of the spectrum will remain.  Symmetries in
the spectra occur when flux is assigned in a way that preserves
the lattice symmetries, and the substitution (\ref{LatSub}) cannot
alter the symmetry properties of the state.

Finally, it is interesting to note that on the lattices with
inversion-symmetric unit cells (namely lattices related to
representations of $SO(N)$ or $SP(2N)$), the operator $G$ which
anti-commutes with the linearized Hamiltonian also anti-commutes
with the lattice Hamiltonian.  This happens because $G$ in these
cases is simply the inversion operator multiplied by an
appropriate gauge transformation, and inversion maps every edge to
another edge associated with the same symmetric generator of the
Lie group representation.  Thus $G$ anti-commutes separately with
all of the symmetric generators -- and hence also with the lattice
Hamiltonian $\sum_{\Sigma^+} (E_\alpha+ E_\alpha^{\dag}) \sin(\bk
\cdot \br_\alpha)$.

\subsection{Extensions to Higher Dimensions}

We have already noted that the generalization of our Hamiltonians
to $d \ge 4$ is problematic. For completeness we note here that
the lattice construction described in Section \ref{LatticeSection}
can be applied to the appropriate representations of the Lie
groups discussed above in arbitrary dimension.  Assigning a
hopping of $\pm i$ to each directed edge will result in a
Hamiltonian related to the group generators by Eq.
(\ref{GeneralH}), for which $H(\bk) = -H (-\bk)$. Additionally,
for all of the cubic lattices (SO(N) spinor, $SO(2N)$ vector, and
the defining representation of $Sp(2N)$ ) a matrix $G$ can be
found which anti-commutes with the Hamiltonian, leading to a
time-reversal invariant spectrum.

In general, however, the symmetry operations of the resulting unit
cell make it impossible to assign flux in such a way that the
lattice symmetries are unbroken.  The notable exception is the
$SO(2N+1)$ spinor case with only the short roots, where all fluxes
are $\pi \equiv - \pi$. This gives the $N$ dimensional Dirac
Hamiltonian.

\section{Conclusions and Outlook} \label{Conclusions}

In this paper we constructed a class of lattices inspired by the
root and weight systems of Lie algebras. The lattices had as their
unit cells minuscule representations of the standard Lie groups
which decorated the underlying lattice $L_{2R}$, elements of the
root lattice with even coefficients. We observed that the lattices
could equally well be viewed as decorations of $L_{2R}$ by the
conjugate representation, which shared corners with the original
one. While construction works for any rank $r$ we stuck to $r=2,\
3$ since these were experimentally accessible and because these
allowed an unambiguous assignment of flux on the triangular faces
of the unit cells. Remarkably, they also correspond in many cases
to known lattices like the pyrochlore, kagom\'{e} or checkerboard.
Even our octachlore lattice is a motif in the perovskite
structure.

We find this last aspect enticing, for it hints that it may be
possible to relate more physics on these lattices to the
underlying Lie algebras. Indeed, as we were finishing up this work
we came across recent work by  Arovas \cite{arovas} who constructs
generalized AKLT models on the kagom\'{e} and pyrochlore lattices
which naturally involve local degrees of freedom that live in the
fundamental representations of $SU(3)$ and $SU(4)$ respectively.

However, our own work makes a different connection. We considered
hopping Hamiltonians which, when written in momentum space, were
elements of the Lie algebra, linear in the momentum $\bk$ for
small $\bk$, and obeyed $H(\bk ) = -H(-\bk )$.\footnote{To belabor
this point, we have an entire unit cell represented by a quark
state, while Arovas has a quark state at each site of the unit
cell.} By varying the signs in front of each generator we could
alter the fluxes in the faces of the unit cell. We found Dirac or
Dirac-like spectra at points, lines and even sheets. The locus of
the zeros had strong ties to the directions of the weights or
roots. We could anticipate and thus understand some of them using
ideas from Lie algebras but often were just able to draw attention
to them. It seems very likely that an assault using ideas from Lie
algebras can yield further understanding. To begin with one must
employ a more systematic way to represent weights and roots in
dual bases: simple weights for the former and simple roots for the
latter. One should also use color groups to classify symmetries of
this problem where the triangular faces of the unit cell are
colored with flux $\pm \pi /2$.  Of all the properties associated
with $H$, the determinant seems most likely to yield to group
theoretic methods. It has uniformly proven to be a much simpler
and more symmetric function of the momenta than individual
eigenvalues.

The spectra often had $E \to - E$ symmetry. For self-conjugate
representations we could fully understand this feature and indeed use our
understanding to construct an operator $G$ that anticommuted with the
Hamiltonian and explained this feature.

While such hopping problems typically arise as lattice regulators for
continuum theories or as mean-field theories for quantum spin models, in this
paper we have studied them in their own right. While we did observe that
most of them are candidates for interesting mean-field theories of quantum
Heisenberg models on the same lattices, a fuller investigation of the
fluctuations would be required to establish their value in that setting.

We hope that some readers will be sufficiently intrigued by the connections that we have
sought to establish in this paper to go on and grapple with them on their own.

\section{Acknowledgements}

R. Shankar thanks Professors Greg Moore and Siddhartha Sahi from
Rutgers and Professor Greg Zuckerman from Yale for helpful
discussions, the National Science Foundation for grant
DMR-0354517, and the Princeton Center for Theoretical Physics for
its hospitality during the course of this work.  F. Burnell and S.
L. Sondhi thank Shoibal Chakravarty for a prior collaboration
which inspired this project.  F. Burnell acknowledges the support
of NSERC. S. L. Sondhi would like to acknowledge support from NSF
Grant No. DMR 0213706.

\bibliographystyle{elsart-num}
\bibliography{LieBib}

\begin{thebibliography}{10}
\expandafter\ifx\csname url\endcsname\relax
  \def\url#1{\texttt{#1}}\fi
\expandafter\ifx\csname urlprefix\endcsname\relax\def\urlprefix{URL }\fi

\bibitem{BA}
G.~Baskaran, P.~W. Anderson, Phys. Rev. B 37~(1) (1988) 580--583.

\bibitem{BAZ}
G.~Baskaran, Z.~Zou, P.~W. Anderson, Solid State Commun. 63~(11) (1987) 973--
  976.

\bibitem{AffleckMarston}
J.~B. Marston, I.~Affleck, Phys. Rev. B 39 (1989) 11538.

\bibitem{HermeleAlg}
M.~Hermele, T.~Senthil, M.~P.~A. Fisher, P.~A. Lee, N.~Nagaosa, X.-G. Wen,
  Phys. Rev. B 70 (2004) 214437.

\bibitem{WenPSG}
X.-G. Wen, Phys. Rev.B 65 (2002) 165113.

\bibitem{Hastings}
M.~B. Hasgings, Phys. Rev. B 63 (2000) 014413.

\bibitem{HermeleKagome}
Y.~{Ran}, M.~{Hermele}, P.~A. {Lee}, X.-G. {Wen}, Phys. Rev. Lett. 98 (2007)
  117205.

\bibitem{LaughlinZou}
R.~B. Laughlin, Z.~Zou, Phys. Rev. B 41 (1989) 664.

\bibitem{WWZ}
X.-G. Wen, F.~Wilczek, A.~Zee, Phys. Rev. B 39 (1989) 11413.

\bibitem{kl}
V.~Kalmeyer, R.~B. Laughlin, Phys. Rev. Lett. 59~(18) (1987) 2095--2098.

\bibitem{Dimers}
D.~S. Rokhsar, Phys. Rev. B 42 (1990) 2526.

\bibitem{ReadSachdev}
N.~Read, S.~Sachdev, Phys. Rev. B 42~(7) (1990) 4568--4589.

\bibitem{Georgi}
H.~Georgi, Lie Algebras in Particle Physics, 2nd Edition, HarperCollins, 1999.

\bibitem{GrBook}
J.-Q. Chen, J.~Ping, F.~Wang, Group Representation Theory for Physicists, 2nd
  Edition, World Scientific Publishing Co., 2002.

\bibitem{Hammermesh}
M.~Hammermesh, Group Theory and its applications, Addison Wesley, New York,
  1962.

\bibitem{ColorGroups}
R.~Lifshitz, Rev. Mod. Phys 69 (1997) 1181, 1218.

\bibitem{KogutSusskind}
J.~Kogut, L.~L.~Susskind, Phys. Rev. D 11~(2) (1975) 395--408.

\bibitem{TorquatoStillinger}
S.~Torquato, F.~Stillinger, J. Appl. Phys 102 (2007) 093511.

\bibitem{MarstonZeng}
J.~B. Marston, C.~Zeng, J. Appl. Phys. 69 (1991) 5962.

\bibitem{wip}
F.~J. Burnell, S.~L. Sondhi, in progress.

\bibitem{shoibal-thesis}
S.~Chakravarty, {PhD} thesis, Princeton University (2004).

\bibitem{bcs-toappear}
F.~J. Burnell, S.~Chakravarty, S.~L. Sondhi, to appear.

\bibitem{HermeleTc}
M.~Hermele, T.~Senthil, M.~P.~A. Fisher, Phys. Rev. B 72 (2005) 104404.

\bibitem{LaughlinZou2}
R.~B. Laughlin, Z.~Zou, Phys. Rev. B 42 (1990) 4073.

\bibitem{Bernier}
J.-S. Bernier, C.-H. Chung, Y.~B. Kim, S.~Sachdev, Phys. Rev. B 69~(21) (2004)
  214427.

\bibitem{arovas}
D.~Arovas, cond-mat.str-el/0711.3921.

\end{thebibliography}

 \end{document}